\documentclass[10pt,conference]{IEEEtran}
\IEEEoverridecommandlockouts
\usepackage{cite}
\usepackage{amsmath,amssymb,amsfonts}
\usepackage{algorithmic}
\usepackage{graphicx}
\usepackage{epstopdf}
\usepackage{textcomp}
\usepackage{xcolor}
\usepackage{subfigure} 
\usepackage{float}
\usepackage{booktabs}
\usepackage[linesnumbered,ruled,vlined]{algorithm2e}
\usepackage{setspace}
\usepackage{svg} 
\usepackage{etoolbox}
\usepackage{url}
\usepackage{newtxtext,newtxmath}
\usepackage{tikz}
\usepackage{tabularx}
\usepackage{booktabs}
\usepackage{multirow}
\usepackage{makecell}
\usepackage{array}
\usepackage{colortbl}
\usepackage[normalem]{ulem}

\newrobustcmd*\blacka[1]{\tikz[baseline=(char.base)]{
            \node[shape=circle,draw,inner sep=1pt,fill,text=white,minimum size=1em] (char) {\textsf{\small a}};}}

\newrobustcmd*\blackb[1]{\tikz[baseline=(char.base)]{
            \node[shape=circle,draw,inner sep=1pt,fill,text=white,minimum size=1em] (char) {\textsf{\small b}};}}

\newrobustcmd*\blackc[1]{\tikz[baseline=(char.base)]{
            \node[shape=circle,draw,inner sep=1pt,fill,text=white,minimum size=1em] (char) {\textsf{\small c}};}}

\newrobustcmd*\blackd[1]{\tikz[baseline=(char.base)]{
            \node[shape=circle,draw,inner sep=1pt,fill,text=white,minimum size=1em] (char) {\textsf{\small d}};}}

\newrobustcmd*\blacke[1]{\tikz[baseline=(char.base)]{
            \node[shape=circle,draw,inner sep=1pt,fill,text=white,minimum size=1em] (char) {\textsf{\small e}};}}

\newrobustcmd*\blackf[1]{\tikz[baseline=(char.base)]{
            \node[shape=circle,draw,inner sep=1pt,fill,text=white,minimum size=1em] (char) {\textsf{\small f}};}}

\newrobustcmd*\blackg[1]{\tikz[baseline=(char.base)]{
            \node[shape=circle,draw,inner sep=1pt,fill,text=white,minimum size=1em] (char) {\textsf{\small g}};}}

\newcommand\name{\text{MESC}}

\begin{document}
\bibliographystyle{elsarticle-num}

\title{MESC: Re-thinking Algorithmic Priority and/or Criticality Inversions for Heterogeneous MCSs}

\author{
    Jiapeng Guan\IEEEauthorrefmark{2}, Ran Wei\IEEEauthorrefmark{3}\IEEEauthorrefmark{2}\IEEEauthorrefmark{1}, Dean You\IEEEauthorrefmark{4}, Yingquan Wang\IEEEauthorrefmark{2}, Ruizhe Yang\IEEEauthorrefmark{2}, Hui Wang\IEEEauthorrefmark{4}, Zhe Jiang\IEEEauthorrefmark{4}\IEEEauthorrefmark{1}\\
    \IEEEauthorblockA{\IEEEauthorrefmark{2}Dalian University of Technology, China.
    \IEEEauthorrefmark{3}Lancaster University, UK.
    \IEEEauthorrefmark{4}Southeast University, China. }
    \thanks{\IEEEauthorrefmark{1} represents corresponding authors. Emails: r.wei5@lancaster.ac.uk and zhejiang.uk@gmail.com.}
}

\newcommand{\zhenote}[1]{\textcolor{red}{Zhe: #1}}

\newcommand{\guan}[1]{\textcolor{blue}{Guan: #1}}

\newcommand{\rt}{\mbox{\sc rt}}
\newcommand{\gpu}{Gemmini$^{\rt}$}
\newcommand{\lo}{\mbox{\sc lo}}

\newcommand{\loFN}{{\mbox{\scriptsize\sc lo}}}
\newcommand{\hi}{\mbox{\sc hi}}
\newcommand{\hiFN}{{\mbox{\scriptsize\sc hi}}}

\newcommand{\gFN}{{\mbox{\scriptsize\sc G}}}
\newcommand{\aFN}{{\mbox{\scriptsize\sc A}}}
\newcommand{\cFN}{{\mbox{\scriptsize\sc C}}}
\newcommand{\rFN}{{\mbox{\scriptsize\sc R}}}
\newcommand{\sFN}{{\mbox{\scriptsize\sc S}}}

\newcommand{\msp}{\mbox{\sc mst}}
\newcommand{\modell}{{\mbox{MESC}}}
\newcommand{\model}{\text{MESC}}
\newcommand{\gpumodel}{Gemmini$^{\rt}$}

\newcommand{\del}[1]{\textcolor{red}{\sout{#1}}}
\newcommand{\add}[1]{\textcolor{blue}{#1}}
\newcommand{\highlight}[1]{\textcolor{black}{#1}}


\maketitle
\IEEEpeerreviewmaketitle

\begin{abstract}
Modern Mixed-Criticality Systems (MCSs) rely on hardware heterogeneity to satisfy ever-increasing computational demands.
However, most of the heterogeneous co-processors are designed to achieve high throughput, with their micro-architectures executing the workloads in a streaming manner.
This streaming execution is often non-preemptive or limited-preemptive, preventing tasks' prioritisation based on their importance and resulting in frequent occurrences of algorithmic priority and/or criticality inversions.
Such problems present a significant barrier to guaranteeing the systems' real-time predictability, especially when co-processors dominate the execution of the workloads (e.g., DNNs and transformers). 

In contrast to existing works that typically enable coarse-grained context switch by splitting the workloads/algorithms, we demonstrate a method that provides fine-grained context switch on a widely used open-source DNN accelerator by enabling instruction-level preemption without any workloads/algorithms modifications. 
As a systematic solution, we build a real system, i.e., Make Each Switch Count (MESC), from the SoC and ISA to the OS kernel.
A theoretical model and analysis are also provided for timing guarantees.
Experimental results reveal that, compared to conventional MCSs using non-preemptive DNN accelerators, MESC achieved a 250x and 300x speedup in resolving algorithmic priority and criticality inversions, with less than 5\% overhead.
To our knowledge, this is the first work investigating algorithmic priority and criticality inversions for MCSs at the instruction level.
\end{abstract}

\section{Introduction}
\label{sc:Intro}
Mixed-Criticality Systems (MCSs) are safety-critical systems in which functionalities are developed at multiple assurance/criticality levels and integrated on a shared platform, e.g., System-on-Chip (SoC)~\cite{burns2013mixed,jiang2021bridging,naghavi2021tolerating,hernandez2020selene,hu2021real}.
For instance, in the automotive industry, an Advanced Driver Assistance System (ADAS) may involve various functionalities developed at different criticality levels, with collision avoidance being of high criticality, whereas route-planning may be of lower criticality~\cite{yang2020all,jiang2020re,iso201126262}.

To satisfy the computational demands of diverse functionalities in MCSs, e.g., running workloads on Deep Neural Networks (DNNs) or transformers, hardware vendors have developed SoCs with different architectures~\cite{Jiang_2022a,ottaviano2023virtualization,majumder2019real}, with a high degree of \emph{heterogeneity}. 
That is, the SoCs couple general-purpose CPUs with heterogeneous co-processors to accelerate algorithmic executions~\cite{latotzke2021efficiency,lu2020hardware,fang2022efficient,genc2019gemmini,genc2021gemmini}. 

\begin{figure}[t]
\centering
\subfigure[Scheduling with a non-preemptive DNN accelerator.]{ 
\label{m1} 
\includegraphics[trim=0.5cm 0.62cm 0.65cm 0.62cm, clip, width=0.98\linewidth]{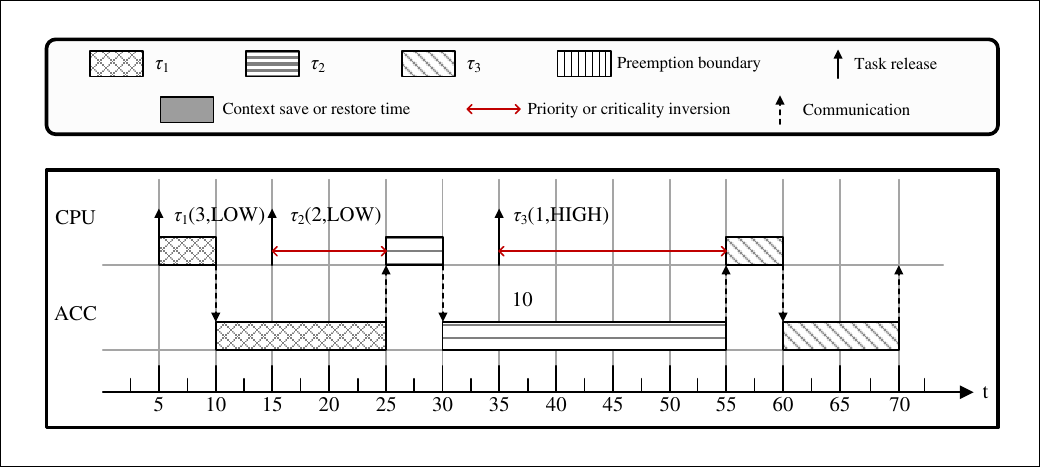} 
}
\subfigure[Scheduling with a DNN accelerator with limited preemption~\cite{liu2020removing,liu2023criticality,liu2023criticality1,liu2024taming,elliott2013gpusync}.]{ 
\label{m2} 
\includegraphics[trim=0.5cm 0.62cm 0.65cm 0.62cm, clip, width=0.98\linewidth]{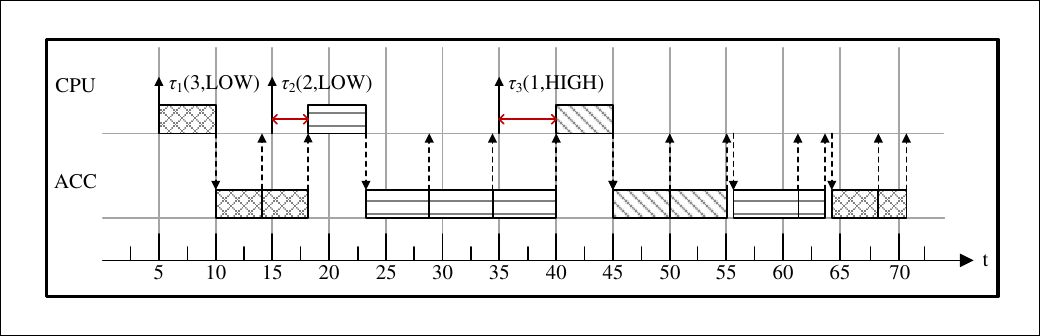} 
}
\subfigure[Scheduling with DNN accelerator allowing instruction-level preemption.]{ 
\label{m3} 
\includegraphics[trim=0.5cm 0.62cm 0.65cm 0.62cm, clip, width=0.98\linewidth]{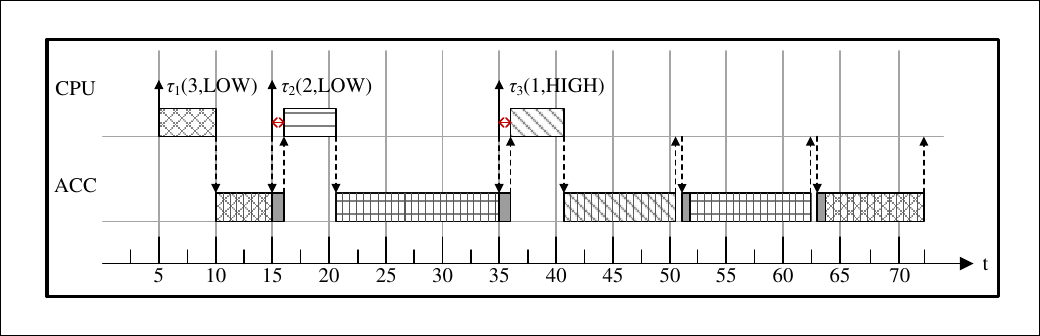} 
}
\caption{Scheduling with DNN accelerators (referred to as ACC in figures) featuring different preemption characteristics: (a) non-preemption, (b) limited preemption, and (c) instruction-level preemption. $\tau_i(P_i, L_i)$ represents task $\tau_i$ with priority $P_i$ and criticality $L_i$, with smaller $P_i$ indicating a higher priority.}
\label{m}
\end{figure}

\noindent \textbf{Research challenges.}
The deployment of these co-processors brings frequent occurrences of \emph{algorithmic priority and/or criticality inversions}\footnote{Due to the differing perspectives on the importance of low-criticality tasks between academia and industry, practical systems cannot tolerate the abandonment of these tasks in degraded modes~\cite{ernst2016mixed,huang2013interference,paulitsch2015mixed,davis2022compensating,zeng2019partition,baruah2011response,burns2013towards}, which may cause criticality inversion. More detailed explanations will be provided in Sec. II.}, posing a significant challenge to guaranteeing system-wide real-time performance~\cite{liu2020removing,liu2023criticality,liu2023criticality1,liu2024taming,lee2020generalized,yoon2022timedice}.
Specifically, unlike CPU micro-architecture, which provides fine-grained\footnote{The ability of the OS kernel to trigger a context switch after the commitment of every single instruction, if no critical section is being executed~\cite{gaitan2015predictable}.} context switches, the micro-architecture of the co-processors is typically designed to maximise throughput while running designated parallel and computation-intensive workloads~\cite{zhou2015gpes,latotzke2021efficiency,lu2020hardware,fang2022efficient,genc2019gemmini,genc2021gemmini}.
This often results in the computation being processed in a streaming order, which is either non-preemptive or only preemptive at the boundaries of the algorithm\footnote{The boundary of an algorithm is 
typically defined as the completion of an operator (e.g., Softmax and ReLu) or a part of the input data.}~\cite{liu2020removing,liu2023criticality,liu2023criticality1,liu2024taming,elliott2013gpusync} (i.e., limited preemption).
Such hardware limitations bar the Operating System (OS) from creating appropriate control flow for context switches in co-processors~\cite{suo2021quantifying,jiang2021hiart}.
Hence, without modifications to the co-processor hardware and/or workloads/algorithms to enhance preemption efficiency, the prolonged occupation of shared resources hinders the effective application of traditional scheduling methods, such as fixed-priority~\cite{baruah2013fixed,li2017fixed} and Earliest-Deadline-First~\cite{burns2017response,burns2017survey,socci2013mixed}, in ensuring schedulability by prioritising task execution based on task importance~\cite{zhao2022minimizing,zhang2023energy,chen2022msrp}.
That is, whilst a less important task occupies the co-processor, more important tasks must wait until the end of the co-processor's ongoing computation.

Figs.~\ref{m1} and~\ref{m2}\footnote{Fig. 1 is a simplified example. In practice, co-scheduling between CPUs and co-processors involves a wide range of strategies~\cite{dong2022schedulability,xiang2019pipelined,lin2023online,xu2021co}, yet most of them still encounter challenges related to priority and/or criticality inversions.} provide an example through a heterogeneous MCS with a CPU and a DNN accelerator.
When the accelerator is non-preemptive (Fig.~\ref{m1}), tasks experience significant priority and/or criticality inversions~\cite{wang2021balancing,amert2023work,basaran2012supporting}, as high-priority/criticality tasks have to be postponed until the accelerator completes all ongoing computations.
When the accelerator is designed/configured to allow limited preemption using existing techniques~\cite{liu2020removing,liu2023criticality,liu2023criticality1,liu2024taming,lee2020generalized,yoon2022timedice} (Fig.~\ref{m2}), the capacity of context switches becomes coarse-grained.
However, since preemption is only possible at the boundaries of the executed algorithm, the system still suffers from considerable priority and/or criticality inversions.

\noindent \textbf{Contributions.}
Here, we introduce \textbf{Make Each Switch Count (MESC)}, a heterogeneous MCS framework that minimises algorithmic priority and criticality inversions down to the instruction level (Fig.~\ref{m3}).
To achieve this, we present 

\begin{itemize}
    \item a new DNN accelerator (\gpu) based on the open-source NPU\footnote{A specialised processor designed to accelerate machine learning and deep learning tasks.} architecture~\cite{genc2019gemmini,genc2021gemmini}, which enables instruction-level preemption. 
    This lays the foundation for fine-grained context switches in heterogeneous MCSs;

     \item a context switch strategy and  OS add-ons to ensure data consistency and manage DNN accelerator context switches. This provides software-level support for DNN accelerator context switches;
     
    \item a full-stack framework that incorporates customised SoC and Instruction Set Architecture (ISA) to OS kernel, forming a complete solution for heterogeneous MCSs;
    
    \item a theoretical model and analysis for the proposed framework provide theoretical validation for \name.
\end{itemize}

We deployed our proposed system on AMD Alveo U280 FPGA and examined it using various metrics, including blocking duration, real-time performance, and overhead.
Experiments show that compared to conventional non-preemptive DNN accelerators, \gpumodel\ achieved 250x and 300x accelerations in resolving algorithmic priority inversion and criticality inversion, respectively.
Furthermore, deploying the \gpu\ in a heterogeneous SoC can significantly improve the timing
performance of MCSs with negligible hardware overhead.

\section{Preliminaries}
\label{sec:preliminaries}

\subsection{Dual-mode MCS}
Conventional MCS theoretical models often assume that the Worst-Case Execution Time (WCET) of a task is estimated with varying degrees of confidence~\cite{vestal2007preemptive, Jiang_2022a, baruah2012preemptive}.
A task's high-critical WCET (\hi-WCET) is associated with a high level of confidence but tends to be overly pessimistic, whereas its low-critical WCET (\lo-WCET) is less pessimistic but has a lower confidence level. 
The correctness criteria specify that if all tasks finish execution within their \lo-WCETs, then they will all finish execution by their deadlines. 
However, if any high-critical task's (\hi-task's) execution time exceeds its \lo-WCET, the \hi-tasks must complete execution by their deadlines~\cite{vestal2007preemptive,guo2018uniprocessor,jiang2020pythia}. 
To satisfy the above criteria, \emph{mode switch} is a straightforward strategy: initially, the system operates in low-critical mode (\lo-mode), in which the system operates under the assumption that the execution time for all tasks will not exceed their \lo-WCETs. 
If this assumption is violated (i.e. if any task fails to finish before its \lo-WCET), the system \textit{switches} to  high-critical mode (\hi-mode). 
In \hi-mode, the scheduling policy may permit \hi-tasks to execute beyond their \lo-WCETs but will ensure that their execution times do not exceed their \hi-WCETs. 
Consequently, to ensure that \hi-tasks meet their deadlines, low-critical tasks (\lo-tasks) may need to be terminated or executed with minimal time budget.
The direct termination of \lo-tasks could lead to potential safety hazards, as the system may still rely on \lo-tasks to perform as expected when entering \hi-mode~\cite{ernst2016mixed,huang2013interference,paulitsch2015mixed,davis2022compensating,zeng2019partition,baruah2011response,burns2013towards}. Thus, similar to Imprecise MCSs~\cite{jiang2021hiart,jiang2022high}, we do not drop \lo-tasks, 
instead, we continue to execute \lo-tasks in \hi-mode, but only when none of the \hi-tasks are being executed.

\subsection{The Gemmini NPU Architecture}

Gemmini~\cite{genc2021gemmini}  is part of the open-source RISC-V ecosystem~\cite{amid2020chipyard}, developed for machine learning by UC Berkeley.
The key architecture is based on a systolic array, including multiple tiles.
Each tile consists of a configurable number of Processing Elements (PEs). 
The array reads data from a local, explicitly managed scratchpad of banked SRAMs, and writes results or intermediate values to a local accumulator.

\noindent \textbf{Gemmini instructions.} 
Gemmini is designed to operate in conjunction with a RISC-V CPU core by executing bespoke, well-defined instructions.
These instructions are dispatched by the CPU and initially received by Gemmini's central controller, forwarded to a reservation station for classification. 
The reservation station categorises the instructions into configuration, load, store and compute. 
Configuration instructions are directly executed, while others are issued to their respective controllers. 
These controllers decompose the instructions and delegate them to different functional modules for execution.

\begin{figure*}[]
\centering
\subfigure[Execution cycles of different workloads.]{ 
\label{fig:sub4}
\includegraphics[trim=1.5cm 9.2cm 2.6cm 5cm, clip,scale=0.29]{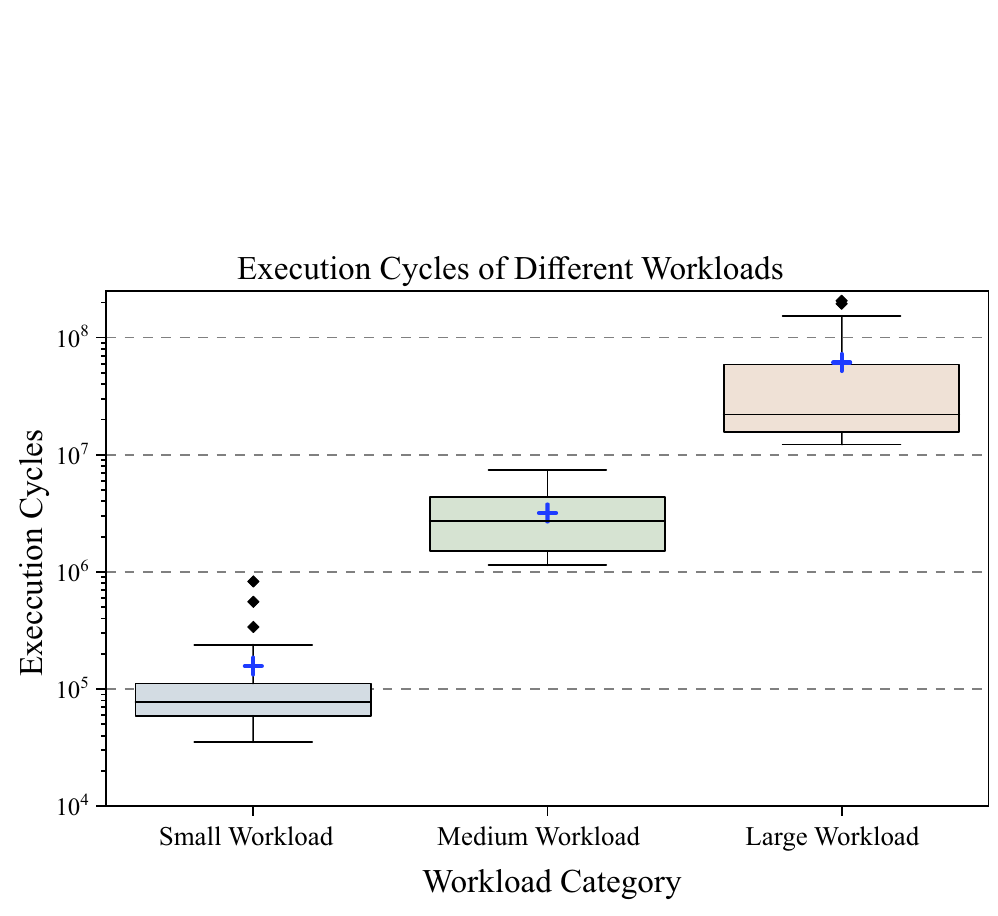}
}
\hspace{10pt} \subfigure[Execution cycles under operator level.]{ 
\label{fig:sub1} 
\includegraphics[trim=1.5cm 9.2cm 2.6cm 5cm, clip,scale=0.29]{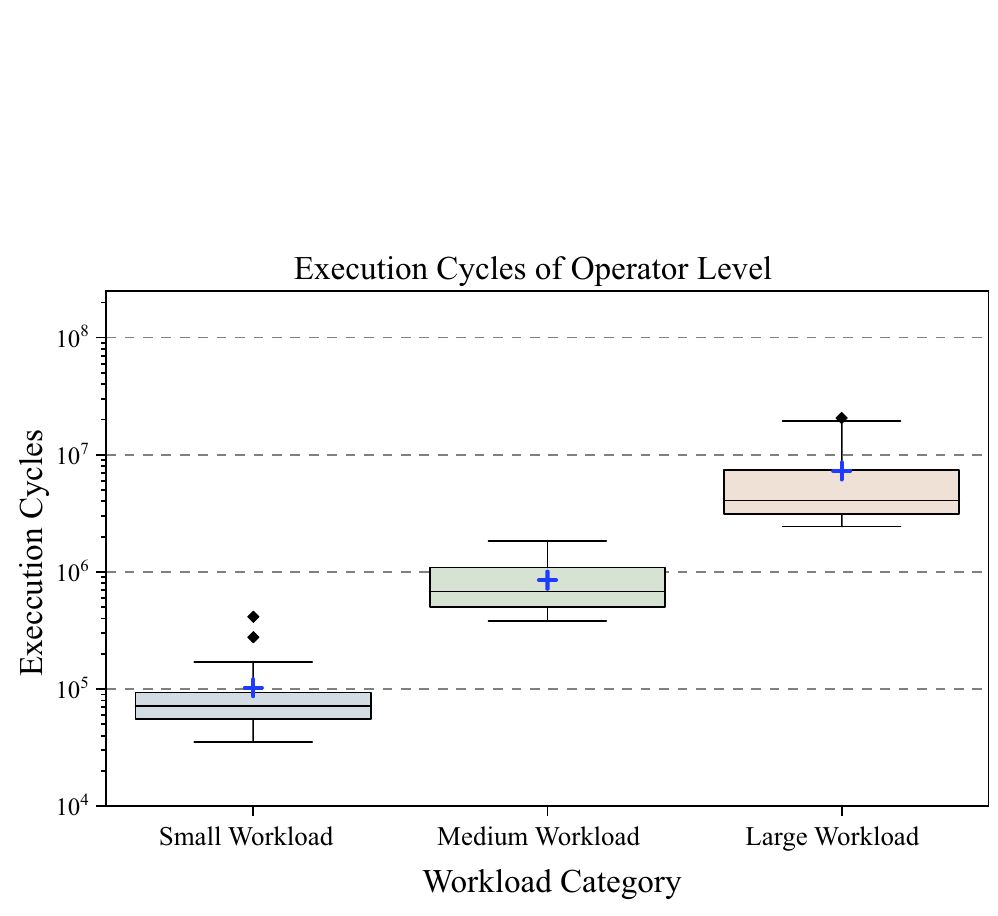} 
}
\hspace{10pt} \subfigure[Execution cycles of insts. in different workloads.]{ 
\label{fig:sub2} 
\includegraphics[trim=0 0 0 0, clip,scale=0.28]{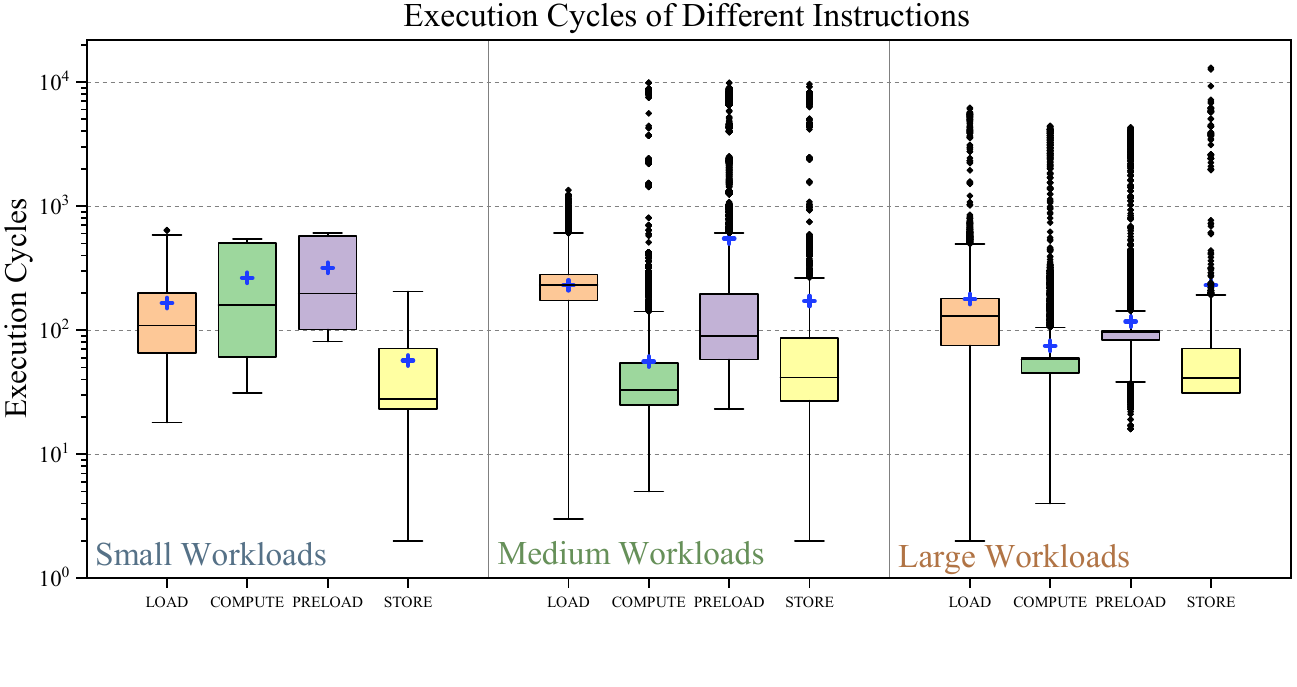}
}
\caption{Execution cycles of Gemmini running workloads of varying sizes.
The workloads are categorised based on their execution times. Small workload: $[0, 1~\text{million}]$ cycles;
medium workloads: $(1~\text{million}, 10~\text{million}]$ cycles;
large workloads: $(10~\text{million}, 1~\text{billion}]$ cycles.}
\label{fig2}
\end{figure*}


\noindent \textbf{Other co-processors.}
As discussed in Sec.~\ref{sc:Intro}, algorithmic priority and criticality inversions are common problems for the MCSs built on heterogeneous SoCs.
Here, we choose Gemmini as the co-processor as our case study for two reasons: 
(i) from the micro-architecture perspective, Gemmini features a representative design similar to other NPUs~\cite{gao2017tetris,qin2020sigma,shin2018dnpu}, facing the same challenges we identified in this work;
(ii) from the SoCs/systems perspective, Gemmini is a popular DNN accelerator that has been integrated into various SoCs~\cite{gookyi2023deep,peccia2022integration,shafique2023csa,vieira2023gem5} and supports modern machine-learning algorithms (e.g., DNNs and transformer), which is crucial for the community.

\section{Motivations and Challenges}
\label{sec:motivation}

In this section, we demonstrate the need for fine-grained context switch in heterogeneous MCSs through the quantitative illustration of the penalty due to algorithmic priority and/or criticality inversions.
We first examine the blocking duration caused by DNN accelerators with different preemptive capacities, then discuss our rationale for selecting MCS as the subject of our studies, followed by an exploration of the challenges in building \name.

\subsection{Quantitative Analysis of Priority and Criticality Inversions}
\label{sbsc:QA}
To quantitatively assess the blocking duration of priority and criticality inversions potentially introduced by DNN accelerators with different preemptive capacities, we execute a collection of workloads of different sizes, using DNNs including AlexNet, MobileNet, ResNet50 and Transformer. 
We developed a cycle-accurate performance counter and integrated it into the instruction commit stage, driven by the same clock source as Gemmini.
This allows us to measure and record the cycles for entire workloads, individual operators (i.e., independent and executable computation blocks or linear operators, such as Softmax and ReLu, within the workload), and single instructions (as illustrated in Fig.~\ref{fig2}), thus reflecting the impact of varying preemption capabilities of different DNN accelerators on the blocking duration of high-priority/criticality tasks.

\noindent \textbf{Non-preemption.}
When the DNN accelerator is non-preemptive, tasks may endure algorithmic priority and/or criticality inversions, with durations that span from tens of thousands of cycles, to several hundred million cycles, as shown in Fig.~\ref{fig:sub4}. 
In safety-critical applications, such prolonged ``obligatory waiting periods'' may cause delays to critical functions that lead to hazardous events, consequently causing accidents that cause harm. 

\noindent \textbf{Limited preemption.}
When the DNN accelerator is featured with limited preemptive capacity, the ``obligatory waiting periods'' can be shortened.
However, the remaining execution time of several million cycles still poses a significant challenge for practical applications, as shown in  Fig.~\ref{fig:sub1}.

\noindent \textbf{Instruction-level preemption.}
When the DNN accelerator enables preemption at the instruction level (as a result of this work, detailed in Sec.~\ref{sc:EE}), the longest execution time for instructions is 2 orders of magnitude shorter than the ``obligatory waiting periods'' in DNN accelerator with limited context switch (even in the worst-case scenario). As shown in Fig.~\ref{fig:sub2}, we classified the instructions into four types\footnote{The preload instruction~\cite{genc2019gemmini,genc2021gemmini} is used to prefetch data into the internal memory (e.g., scratchpad and accumulator), working in conjunction with compute instructions to optimise data transfer and pipeline execution.} and plotted their execution cycles.

Hence, with the provision of instruction-level switches, the duration of priority and criticality inversions could be reduced to the time needed for instruction and context switch, allowing timely preemption for high-priority/criticality tasks.

\begin{figure*}[t]
\centering
\includegraphics[trim=0.5cm 0.62cm 0.65cm 0.68cm, clip, width=0.95\textwidth]{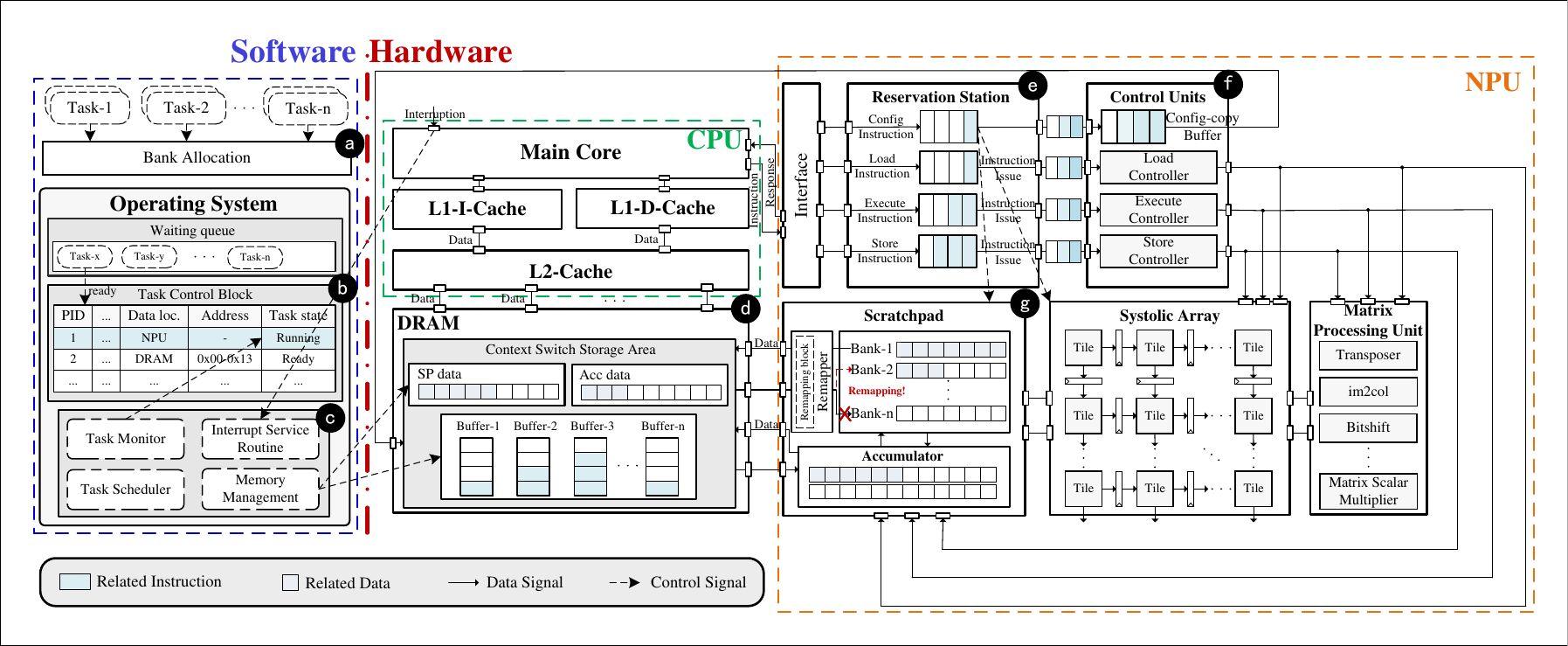}
\caption{Architectural overview: the blue box represents the software level, the orange and green boxes correspond to the NPU and CPU (hardware), respectively.}
\label{f5}
\end{figure*}

\subsection{MCS: an Example}

Although contributions to priority inversion issues are applicable to a wide range of real-time systems, we have chosen MCS as our system model for the following reasons.

\noindent \textbf{System complexity.}
In the context of algorithmic priority inversion, MCS introduces an additional challenge: criticality inversion~\cite{ernst2016mixed,huang2013interference,paulitsch2015mixed,davis2022compensating,zeng2019partition,baruah2011response,burns2013towards,jiang2021hiart,jiang2022high,zou2023context}. 
In this regard, MCS is not only a typical complex system but also a representative of systems with broader real-time system challenges (i.e., other real-time systems can be viewed as subsets of MCS).
By addressing both the broader issue of priority inversion in real-time systems and the more specific problem of criticality inversion in MCS, we demonstrate the effectiveness of our approach, without limiting its applicability to other systems.

\noindent \textbf{Popularity.}
MCS has remained a vital and popular research area for two decades, attracting continuous contributions and advancements from the research community~\cite{jiang2019mcs,jiang2020pythia,jiang2020re,jiang2021bridging,jiang2021hiart,jiang2022pspsys,burns2020schedulability,burns2022approach,davis2022compensating}, and playing a crucial role in industries with high safety-critical requirements~\cite{burns2013mixed,jiang2021bridging,naghavi2021tolerating,hernandez2020selene,hu2021real} (as detailed in Sec. I).

\subsection{Research Challenges}
Based on our observations, it is crucial to enable instruction-level context switches for DNN accelerators to minimise algorithmic priority and criticality inversions.
However, achieving this is not straightforward, as several key challenges must be addressed:

\begin{itemize}
\item \textbf{Complexity}. It is important to overcome the inherent complexity involved in DNN accelerator context switch, to properly handle the saving and loading of computation and configuration data to achieve comprehensive accelerator context switches (and develop an acceleration strategy);

\item \textbf{Scheduling}. The OS scheduler  shall be enhanced to orchestrate context switch in the accelerator to correctly schedule tasks and provide software-level support for context switch so that the accelerator switch becomes transparent.

\item \textbf{Hardware and software integration}. System-level integration of hardware and software is needed for a generic approach in building heterogeneous MCSs with DNN accelerators that support instruction-level context switches.

\item \textbf{Modelling and analysis}. Theoretical modelling of the system is needed for a comprehensive theoretical analysis to show that our proposed approach is fit for purpose.
\end{itemize}

\section{\model: the Systematic Framework}
\label{sec:approach}
In coping with the above challenges, we have developed \name, which includes a DNN accelerator (called \gpu), allowing instruction-level preemption, and OS add-ons that provide software-level support for context switch control flow. 
The architectural overview of \model\ is given in Fig.~\ref{f5}.

\noindent \textbf{Hardware level.} 
At the hardware level, we have developed a context switch and acceleration mechanism for the DNN accelerator. 
During the context-saving process, computation data is directly transferred from the scratchpad to the DRAM (Fig.~\ref{f5}.\blackd\ ), and the address information is recorded in the task's Task Control Block (TCB, Fig.~\ref{f5}.\blackb\ ). 
For configuration data, we employ a config-copy buffer (Fig.~\ref{f5}.\blackf\ ) to store the most recently executed configuration instructions of different types, and these are also sent to the DRAM (Fig.~\ref{f5}.\blackd\ ). 
During the context-restoring process, computation data is re-loaded into the local memory of the accelerator (Fig.~\ref{f5}.\blackg\ ) using addresses recorded in the TCB. 
The accelerator will also be reconfigured by the stored configuration instructions. 
Instructions previously dispatched by the CPU without receiving a response signal from the accelerator are then resent to it.
Also, we implemented an address remapper (Fig.~\ref{f5}.\blackg\ ), which centralises the storage of data into allocated banks and uses a remapping block to maintain this process, thereby supporting the subsequent local memory\footnote{The ``local memory" allocated by this method specifically refers to the scratchpad in this paper.} allocation methods we employ.
The remapping block is sent back to the DRAM during each context switch and updated during each context restoration, based on the bank addresses into which the computation data is reloaded.
When there is sufficient local memory, a minimal amount of computation data needs to be preserved, reducing the time required for context switches.
Lastly, we have added specific ISA to the system, (Tbl. I), which we will use later in Secs.~\ref{sec:hardware} and ~\ref{sec:software}.

\begin{table}[t]
\centering
\caption{New ISA added for \gpumodel.}
\hspace{100pt}\resizebox{.95\columnwidth}{!}{%
\begin{tabular}{c|l}
\hline
\rowcolor[HTML]{EFEFEF} 
\textbf{Name/Type}                                                               & \multicolumn{1}{c}{\cellcolor[HTML]{EFEFEF}\textbf{Description}}                                                                                                          \\ \hline
\textbf{instruction\_freeze}                                                     & \begin{tabular}[c]{@{}l@{}}Halts the execution of all queued insts., except \\ for flush-related insts.\end{tabular}                                                       \\ \hline
\textbf{\begin{tabular}[c]{@{}c@{}}step\_wise\_mvin\\ /mvout\end{tabular}}       & \begin{tabular}[c]{@{}l@{}}Moves data into/out of the scratchpad using \\ the configuration channel.\end{tabular}                                                         \\ \hline
\textbf{mvin/mvout\_config\_buffer}                                                   & Moves stored configuration insts. from/to DRAMs.                                                                                                                               \\ \hline
\textbf{reconfig}                                                   & Executes and clears insts. in the config-copy buffer.                                                                                                                               \\ \hline
\textbf{\begin{tabular}[c]{@{}c@{}}mvin/mvout\_remapping\\ \_block\end{tabular}} & \begin{tabular}[c]{@{}l@{}}Moves the remapping block into/out of the accelerator\\ using the default configuration channel.\end{tabular}                                          \\ \hline
\textbf{flush\_x}                                                                & \begin{tabular}[c]{@{}l@{}}The x represents freeze, bank, etc., either \\ resuming the execution of insts. or flushing data \\ within a specified component.\end{tabular} \\ \hline
\end{tabular}
}
\end{table}

\noindent \textbf{Software level.} 
At the software level, we provide an accelerator local memory allocation method (Fig.~\ref{f5}.\blacka\ ), along with a task monitor and scheduler (Fig.~\ref{f5}.\blackc\ ) that can be adopted by (real-time) OS kernels.
This method, which is enhanced by the address remapper (Fig.~\ref{f5}.\blackg{}), adjusts the local memory allocated to tasks, thereby directly implementing hardware segmentation to reduce the time required for context switches when local memory is sufficient. This is  under the fundamental premise of maintaining unchanged task execution times.
We also added the task monitor and scheduler to the OS kernel space. 
The monitor uses hardware timers to track the running conditions of tasks, while the scheduler provides control functions, e.g., \texttt{Context\_switch} and \texttt{Mode\_switch} to manage tasks and mode switches.
In the OS user space, we preserve compatibility with existing OS interfaces, enabling the migration and adaptation of tasks initially designed for traditional MCS.

\noindent \textbf{Context switch.} 
A context switch is often triggered by an interrupt or a system call, where the OS transitions to the Interrupt Service Routine (ISR) to manage the current situation. 
If the interrupt is triggered by a timer monitoring the task (other interrupts are handled by the original OS kernel), control reverts to the task scheduler (Fig.~\ref{f5}.\blackc\ ).
The scheduler then determines whether a context switch is necessary.
If a context switch is required (detailed in Sec. VI), it initially prohibits the execution of instructions that are queued in the accelerator except \texttt{flush} instructions and waits for the completion of multiple instructions that are concurrently executing (Fig.~\ref{f5}.\blacka{} and \blackf\ ).
The scheduler then flushes the accelerator instruction queues (Fig.~\ref{f5}.\blacke{} and \blackf{}), ensuring no instructions are present in the accelerator except config-copy buffer. It reinstates the issuing and operation of instructions in the accelerator to ensure that the instructions related to context switch are not impeded. 
It then directs the accelerator to save the current context to DRAM (Fig.~\ref{f5}.\blackd\ ), including the computation and configuration data from the accelerator. 
When a task completes and a previously preempted task needs to resume execution, a similar approach is adopted. The relevant data is retrieved from DRAM and restored to the accelerator, the remapping block is updated, and configuration of the accelerator is returned to its previous state. Subsequently, the OS directs the CPU to re-dispatch the instructions that were previously dispatched to the accelerator but did not receive a response signal. This step restores the instructions that were queued in the accelerator. 
Following this procedure, the accelerator is fully restored to its state prior to the context switch.

\noindent \textbf{Mode switch.} 
The strategy of mode switch in \model\ is governed by the following rules:
\begin{itemize}
\item \textbf{\lo-mode:} the system prioritises the highest-priority tasks for execution. Under the bank allocation method, all tasks are executed using minimal accelerator local memory necessary to maintain their operational speed.
\item \textbf{Mode transition:} when a \hi-task exceeds its \lo-WCET, the system commences mode transition. The system prioritises the scheduling of \hi-tasks, and \lo-tasks are scheduled only when all \hi-tasks are idle.
During transition, only \lo-tasks with computation data not yet saved back to the main memory are permitted to execute until the accelerator local memory contains data from at most one \lo-task.
\item \textbf{\hi-mode:} once the accelerator local memory contains data from at most one \lo-task, the system transitions to \hi-mode. The system prioritises the scheduling of \hi-tasks with accelerator local memory allocated according to the bank allocation method. 
When \lo-tasks preempt each other, all related data must be evacuated from accelerator to ensure that the accelerator local memory contains data from at most one \lo-task at any time, thereby reducing the average criticality inversion duration for \hi-tasks.
Moreover, when there are no tasks currently executing within the system, it will revert to \lo-mode.
\end{itemize}

\noindent \textbf{Framework availability.}
The MSEC system architecture is designed with a high degree of availability.
Specifically, (i) the architecture can be adapted with minor modifications to other DNN accelerators (e.g., Tetris~\cite{gao2017tetris}, Sigma~\cite{qin2020sigma}, and DNPU~\cite{shin2018dnpu}); 
(ii) modifications to the OS retain the Application Programming Interfaces (APIs) used in traditional MCS frameworks, allowing user programs developed for traditional MCS frameworks to be directly ported to this system without major changes; 
(iii) in MESC, the CPU primarily functions as a instruction dispatcher to the accelerator. However, our approach can still be applied to systems employing other co-scheduling strategies to address challenges arising from priority and criticality inversions; 
(iv) although our research focuses on MCS, it is general enough to be applicable to other real-time systems as well;
and (v) the MESC can be ported to other systems without altering the programming model. Specifically, the Gemmini programming model, which includes a high-level model~\cite{genc2019gemmini,genc2021gemmini} for reading DNN descriptions from the Open Neural Network Exchange (ONNX) format and generating software binaries, as well as a low-level model~\cite{genc2019gemmini,genc2021gemmini} for invoking library functions provided by UC Berkeley, remains unchanged.

\section{\gpu: the Micro-architecture}
\label{sec:hardware}

As introduced, we use Gemmini as a case study to demonstrate the efficacy of our framework, leading to the evolution of Gemmini into what we call \gpumodel\ with the integration of a context switch mechanism. 
To do so, we outline three key requirements: \gpumodel\ should (i) maintain the consistency of computation data, (ii) maintain the consistency of configuration data, and (iii) allocate hardware resources efficiently and optimise the context switches' overhead.

In response to these requirements, we designed a novel micro-architecture that includes (i) a default configuration channel that operates independently of the current accelerator configuration environment; (ii) a config-copy buffer that stores the most recent types of configuration instructions; and (iii) an address remapper that assists the accelerator local memory allocation.

\subsection{Handling Computation Data}
In Gemmini, the matrix processing units (e.g., transposer, bit shifter) function as operational modules that terminate upon completion of their instructions and do not store data. Similarly, the systolic array can also be seen as a specialised computational unit. It executes operations based on instructions, storing intermediate results in the accumulator upon completing a single computation~\cite{genc2019gemmini,genc2021gemmini}. Subsequent instructions retrieve values from the accumulator rather than relying on registers of the systolic array. Thus, all computation data is stored exclusively in the scratchpad and the accumulator. 
Data and memory management between the accelerator and the host CPU is explicit in Gemmini, meaning that data must be explicitly moved between the main address space of the processor and the private address space of the accelerator using a series of move instructions. 
In the original ISA, two data movement instructions, \texttt{mvin} and \texttt{mvout}, facilitate the transfer of data between the main memory and the internal accelerator via the Direct Memory Access (DMA) unit.

\begin{figure}[t]
\centering
\includegraphics[trim= 1.1cm 0.5cm 0.5cm 0.65cm, clip, width=0.95\linewidth]{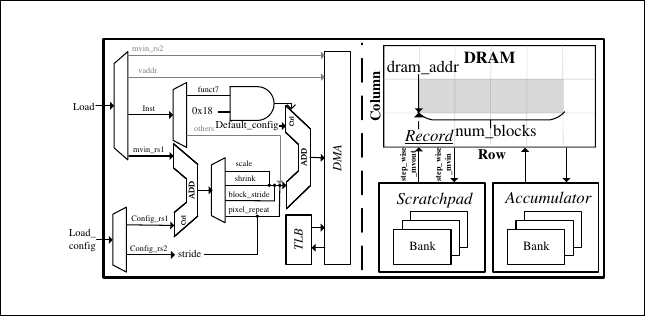}
\caption{The default configuration pathway for load class instructions (shown on the left, in black), and the process utilising \texttt{step\_wise\_mvin} and \texttt{step\_wise\_mvout} to transfer matrices into or out of the scratchpad and accumulator (right side).}
\label{f6}
\end{figure}

However, it is important to note that configuration instructions may influence these move instructions. 
If \texttt{mvin} and \texttt{mvout} are executed on an actively running accelerator, the data saved and restored could be incomplete. 
Therefore, to ensure configuration consistency and to secure the complete retrieval of computation data, we have introduced new move instructions, \texttt{step\_wise\_mvin} and \texttt{step\_wise\_mvout}, as shown on the right of Fig.~\ref{f6}. 
These new instructions utilise the default configuration channel (shown on the left of Fig.~\ref{f6}), which we developed within the control units of the accelerator, enabling the complete saving and restoring of computation data without altering the current accelerator configuration environment. 
Specifically, during the context-saving process, by utilising these move instructions and recording the target main memory addresses before the data transfer, we can ensure the complete preservation of computation data in the main memory through the scheduling of the OS. 
Similarly, during the context-restoring process, by retrieving the main memory address index of the computation data, we can fully load it back into the accelerator.
Through such processes, we can completely save the computation data, and restore it as needed at any time.

\subsection{Handling Configuration Data}
In Gemmini, configuration instructions are unique compared to other types of instructions, for a couple of reasons. 
Firstly, configurations are categorised into four types: load, store, execute, and norm, where a subsequent configuration instruction of the same category can override the previous one. 
Secondly, once a configuration instruction is dispatched by the CPU to the accelerator and enters the reservation station, it does not need to be issued to the execution units. 
Instead, it is executed immediately within the reservation station, directly affecting the corresponding registers to achieve the configuration purpose, with a 2-cycle execution time.

To accommodate the characteristics outlined above, we have implemented a dedicated buffer called the config-copy buffer to store the four most recent configuration instructions of each type. 
During the context-saving process, these instructions are transmitted back to DRAM using the newly added \texttt{mvout\_config\_buffer} instruction, with their addresses recorded by the monitor. 
During the context-restoring process, these instructions are re-loaded into the accelerator using the \texttt{mvin\_config\_buffer} instruction, and executed after all data has been fully reloaded to reconfigure the accelerator.

In scenarios involving complex system configurations, it is feasible to implement and maintain multiple high-level configuration instructions to streamline system setups and manage configurations. 
If such strategies prove insufficient due to overly intricate configurations, an alternative viable method is to introduce a flag bit for cache lines in the L1 I-cache of the CPU. 
Configuration instructions executed during task operations would update these flag bits, and the scheduler would track their locations, ensuring these cache lines are not displaced during cache entry switching. 
When tasks are finished or terminated, the associated cache line flags are cleared. 
During context restoration, these flagged cache lines are prioritised for execution. 
However, as noted in our Gemmini case study, due to the standardised nature and minimal quantity of configuration instructions, we opted not to employ this method.

\subsection{The Address Remapper}
Given that not every task requires all accelerator resources, to enable more efficient resource utilisation, we introduced an address remapper to transition the explicit resource allocation in Gemmini to a semi-explicit form. 
This allows for effective hardware partitioning when local memory is sufficient, eliminating the need to save and restore scratchpad data during context switches and accelerating the context-switching process.

\begin{figure}[t]
\centering
\includegraphics[trim= 0.7cm 0.65cm 0.5cm 0.65cm, clip, width=0.95\linewidth]{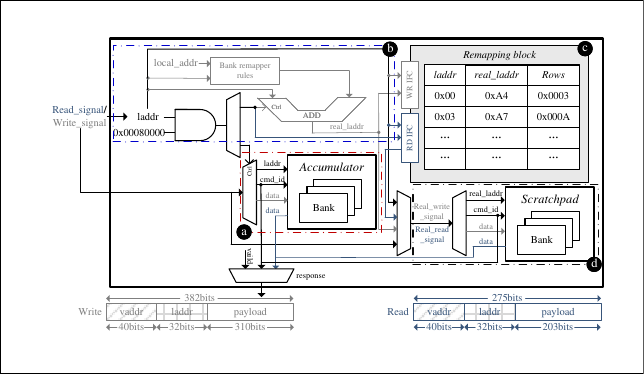}
\caption{Micro-architecture of the address remapper, featuring the signal \texttt{laddr} for the local memory address to be written or read, and \texttt{local\_addr} representing the address calculation rules within the accelerator. Blue lines: the path of the read signals; grey lines: the path of the write signals; black lines: the common pathways.}
\label{f7}
\end{figure}

The address remapper, as shown in Fig.~\ref{f7}, operates by intercepting the DMA streams entering the scratchpad, incorporating a dynamic offset to relocate them to the appropriate bank (Fig.~\ref{f7}.\blackb{} and \blackd{}). 
This ensures that, externally, it appears as though the data is still being directed to its initial position within the scratchpad.
We established a configurable-sized remapping block, defaulting to 4KB, within the scratchpad  (Fig.~\ref{f7}.\blackc{}) that stores the mapping from their original addresses to actual addresses, facilitating address translation and data management. 

Specifically, we have assigned a semaphore to each bank, referred to as the banklock, which is engaged when the bank is loaded with valid data. 
In the event of a task anomaly or task completion, during the context-switching process, the OS identifies which bank is involved by consulting the address mappings in the remapping block. 
It then deactivates the corresponding semaphore and flushes the data contained within that bank.
When the DMA stream initiates a write request to the scratchpad, the system intercepts the stream and applies a dynamic offset to the scratchpad address bits before the stream enters the write queue (Fig.~\ref{f7}.\blackb{}). 
This offset positions the data into a scratchpad bank that is either partially filled and locked by the task, or into one that is currently unlocked. 
The system then locks the bank and records the start and end positions of the address in the remapping block (Fig.~\ref{f7}.\blackc{}). 
It is crucial, however, to ensure that the number of scratchpad banks that are accessed does not exceed the total number allocated by the system.
Similarly, when the DMA stream requests a read from the scratchpad, the system first consults the remapping block (Fig.~\ref{f7}.\blackc{}) to locate the corresponding mapping and then adjusts the scratchpad address bits (Fig.~\ref{f7}.\blackd{}) in the DMA stream. 
When a task is preempted and subsequently resumed, the remapping block will adjust the target address based on the newly allocated bank that the task enters. 
In fact, in \gpumodel, the accumulator is almost identically structured to the scratchpad. However, compared to the scratchpad, the memory capacity of the accumulator is significantly smaller. 
Based on this, we have chosen not to impose allocation restrictions on the accelerator’s accumulator (Fig.~\ref{f7}.\blacka{}).

Under this mechanism, the OS only needs to keep track of the number of banks currently allocated to schedule tasks effectively. 
The specific banks and addresses allocated are determined automatically by the hardware, ensuring efficient resource management and task execution.

\section{The OS Kernel Add-ons}
\label{sec:software}
To ensure compatibility with the hardware mechanisms for context switches, we developed two software add-ons for real-time OS kernels (e.g., FreeRTOS) running on an open-source RISC-V processor (e.g., Rocket core~\cite{asanovic2016rocket}): (i) a task monitor to oversee task properties (including data locations) and key system checkpoints, 
and (ii) a task scheduler 
to help manage mode switches and context switches.

\subsection{The Task Monitor}
When an interrupt occurs, the ISR calls the task monitor to add or update task information within the TCB, storing task information, i.e., basic attributes, a program counter, data storage locations (DRAM or accelerator local memory) along with their addresses, and their status. 
The storage location of the task data determines if the computation data needs to be reloaded into the scratchpad during context restoration. 
A task's current status can be: ready, running, pending, or interrupted.

\begin{algorithm}[t]
\DontPrintSemicolon
\SetAlgoLined
\small
\label{al1}
\tcc{Kernel.Scheduler:Context Switch}
\SetKwFunction{FMain}{Context\_switch}
\SetKwFunction{FContextSave}{Context\_save}
\SetKwFunction{FContextRestore}{Context\_restore}
\SetKwProg{Fn}{Function}{:}{End Function}
    \Fn{\FMain{task *current}}{
    task *next = NULL\;
    \textbf{Kernel}.Intr.Disable()\;
    next = \textbf{Kernel}.Scheduler.Find\_next\_task()\;
    \eIf{(current == NULL)}{
        continue;
    }
    {
        \eIf{(current == next)}{
            continue\; 
        }{
        \textbf{ACC}.Instructions\_freeze()\;
        \While{(\textbf{Kernel}.Instructions\_complete() == False)}{
            continue\; 
        }
        \textbf{Monitor}.Timer.Pause(current)\; 
        \textbf{Kernel}.Scheduler.\FContextSave(current, next)\;} 
        }

    \If{(next != NULL \textbf{and} next != current)}{
        \eIf{(\textbf{Monitor}.Timer.Is\_zero(next))}{ 
        \textbf{Monitor}.Timer.Set(next)\;} 
    {
        \textbf{Kernel}.Scheduler.\FContextRestore(current, next)\; 
    }
    \textbf{Monitor}.Timer.Activate(next)\; 
    current = next\;
    }
    \textbf{Kernel}.Intr.Enable()\; 
    \textbf{Kernel}.Scheduler.Context\_jump\_to\_PC(current)\; 
}
\tcc{Kernel.Scheduler:Context Save}
\Fn{\FContextSave{task *current, task *next}}{
    \textbf{ACC}.Preprocess\_handler()\; 
    \textbf{ACC}.Accumulator\_step\_wise\_mvout()\;
    \textbf{ACC}.Mvout\_config\_buffer()\;
    \If{(next-\textgreater Bank + \textbf{Kernel}.Scheduler.Locked\_banks() $\leq$ \textbf{Kernel}.Scheduler.Total\_banks())}{
        \textbf{ACC}.Banks\_step\_wise\_mvout()\;
        \textbf{ACC}.Release\_banks(current)\;
    }
    \textbf{ACC}.Flush()\;
}
\caption{Function for Context Switch} 
\end{algorithm}

Similar to traditional dual-criticality MCS frameworks, the task monitor creates a timer for each task. 
Specifically, upon system initialisation, the \lo-WCETs of tasks are preloaded into memory. During context switches, the monitor suspends the current task's timer and activates the next task's timer. 
If a \hi-task exceeds its \lo-WCET in \lo-mode, the timer will trigger an interrupt to execute the \texttt{Mode\_switch} function to transition the system mode. 
Upon task completion, this timer is reset.

\subsection{The Task Scheduler}


The task scheduler follows the system rules of \model, as outlined in Sec.~\ref{sec:approach}. 
Its responsibilities include:
(i) the scheduling of an idle task when no other tasks are present, continuously checking for tasks that require execution;
(ii) the routine maintenance of system status and assessing the necessity for context switches;
(iii) the invocation of the \texttt{Context\_switch} function and its auxiliary functions to assist the accelerator in performing context switches;
and (iv) upon completion of the current task, flushing its data in the accelerator and identification of the next task that requires execution.

During task execution, the scheduler periodically checks for arriving tasks and maintains the system status using a hardware timer. 
The pseudocode in Alg.~\ref{al1} details the steps involved for context switches.
Initially, the kernel disables interrupts and the scheduler identifies the next task to execute.
The scheduler then evaluates how to manage the context based on the current system task status. Specifically, if no tasks are executing, there is no need to save the context (Alg.~\ref{al1}: lines 5 - 6). 
If a task is executing, the scheduler checks whether it is the same as the current task. 
If they are the same, no further action is taken and interrupts are re-enabled before returning (Alg.~\ref{al1}: lines 8 - 9, 28 - 30). 
In other scenarios where accelerator context needs to be saved, the scheduler first prohibits the execution of instructions in the accelerator except for \texttt{flush} instructions which are neither issued in the reservation station nor started in the control module (Alg.~\ref{al1}: line 11). 
The kernel waits for response signals to ensure that all executing instructions are complete (Alg.~\ref{al1}: lines 12 - 14). 
Subsequently, the task monitor pauses the timer of the current task. 
The scheduler invokes the \texttt{Context\_save} function to save the current context.

In \texttt{Context\_save}, preprocessing is initially performed. Specifically, all instruction queues within the accelerator, except for the config-copy buffer, are flushed, and the issuing and execution of instructions are resumed.
Computation data and configuration instructions are then preserved (Alg.~\ref{al1}: lines 33 - 34). 
The handling of data within the scratchpad is then determined by whether the number of banks in the scratchpad is sufficient to support the normal operation of subsequent tasks (Alg.~\ref{al1}: lines 35 - 38). 
Finally, the accelerator flushes all related data, including instructions in the queue, Translation Lookaside Buffer (TLB), and configuration data.

After the current context has been processed, if a task needs execution and data restoration, the \texttt{Context\_restore} function is invoked (Alg.~\ref{al1}: lines 19 - 24). 
\texttt{Context\_restore}'s logic mirrors \texttt{Context\_save}. 
It decides whether to re-load task data based on the stored addresses, updates the remapping block, re-configures the accelerator as necessary, and re-sends instructions that did not receive a response signal previously.
Following context restoration, the monitor activates the timer for the task. 
Finally, the kernel enables interrupts and resumes task execution. 
It is important to note that this discussion only addresses scenarios that require a context switch. 
Situations that do not lead to a context switch are managed in the upper-level function of \texttt{Context\_switch} within the scheduler.

Despite the introduction of a new system architecture in \model, the design of the context switch minimises modifications, which are limited to slight adjustments to the TCB and the addition of a task monitor and a task scheduler, to the OS. 
Additionally, this design maintains the original OS APIs used in traditional MCS frameworks. 

\section{Timing Analysis and Optimisation}

In this section, we present the Worst-Case Response Time (WCRT) analysis for \model\ to assess the schedulability of a given task set with constrained deadlines. 
We also propose a local memory allocation method for distributing memory resources within the accelerator to improve memory utilisation and isolate the data of different tasks when local memory is sufficient, thereby reducing the time required for context switches.

\subsection{The Task Model}
In our system, which consists of a CPU and a DNN accelerator, we consider a set of $n$ sporadic tasks, denoted by $\Gamma =\{\tau_1,\dots,\tau_n\}$, and a shared resource (e.g., \gpumodel) which may be required by a task.
On startup, each task is initialised, 
followed by the requisite access of the shared resource until its execution completes. 
Each task $\tau_i$ is characterised by a tuple of parameters 
$(P_i,T_i,D_i,C_i^{\loFN},C_i^{\hiFN},L_i,\eta_i)$.
$P_i$ denotes the priority of task $\tau_i$, which is determined using a fixed priority assignment scheme in the system.
$T_i$ denotes the period or minimum inter-arrival time of task $\tau_i$.
$D_i$ denotes the deadline of task $\tau_i$. 
$C_i^{\loFN}$ denotes the \lo-WCET of task $\tau_i$, and $C_i^{\hiFN}$ denotes its \hi-WCET, with the condition that for any task $C_i^{\loFN} \leq C_i^{\hiFN}$.
$L_i$ denotes the criticality level of task $\tau_i$.
$\eta_i$ denotes the number of scratchpad banks allocated to task $\tau_i$.



\subsection{Worst-case Response Time Analysis}
We now present the WCRT analysis. 
We begin by categorising tasks for a given task $\tau_i$. This categorisation is based on priority and criticality levels, leading to the formation of four categories: 
$hpH(\tau_i)$ and $hpL(\tau_i)$ denote the sets of \hi-tasks and \lo-tasks with higher priority than $\tau_i$;  
$lpH(\tau_i)$ and $lpL(\tau_i)$ denote the set of \hi-tasks and \lo-tasks with lower priority than $\tau_i$.

Blocking occurs if a task is not scheduled when it is supposed to be, classified as either \emph{criticality-inversion blocking (ci-blocking)}, caused by a \lo-task (even a high-priority \lo-task) holding a shared resource needed by a \hi-task in mode transition or \hi-mode; or \emph{priority-inversion blocking (pi-blocking)}, caused by a task holding a resource needed by a higher-priority task.

In \model, we only need to consider the blocking caused by tasks requiring the accelerator, as the context switches for CPU-only tasks are completed in such a short time that they cause negligible blocking. We can initially perform a preliminary task partitioning using a function $F(\Gamma)$, which returns the subset of tasks within $\Gamma$ that require accelerator utilisation. The complement of this subset, representing the tasks that only require the CPU, can be returned using $\overline{F}(\Gamma)$. For the accelerator, we have reduced the blocking impact caused by shared resources to the instruction level. This can be achieved by utilising a function $I(\Gamma)$, which provides the longest execution time of instructions on the accelerator within the task set $\Gamma$. 
The task scheduler operates at intervals of $T_{sr}$ to routinely maintain system status and assess the necessity for context switches. 
We define some notations to describe the time consumption in different scenarios:
$\Upsilon_{\aFN sr}^\sFN$ and $\Upsilon_{\aFN sr}^\rFN$ represent the maximum durations of the combined CPU context switch and accelerator context-saving time when a context save is required, and context-restoring time when a context restore is required, respectively.
$\Upsilon_{\cFN sr}$ denotes the maximum duration of the CPU check time.
$\Upsilon_{\cFN sr}^\cFN$\footnote{The context switch for CPU-only tasks has not been further differentiated into save and restore process because, unlike the accelerator, this duration is significantly shorter, and the time for saving and restoring is similar, lacking the distinct variability observed with accelerator.} denotes the maximum time required for a context switch involving the CPU-only tasks.
All parameters can be determined through experimental measurements. 
In our theoretical analysis, the periodic maintenance tasks performed by the task scheduler under these conditions are considered as the tasks with the highest priority.
Given that these scenarios represent different branches of the same task scheduler behaviour, they will not coexist in the kernel. 
For clarity, we list these scenarios separately and do not categorise them under $hpH(\tau_i)$.

We consider three cases for the schedulability analysis~\cite{Baruah_2011}:
\begin{enumerate}
\item The schedulability of \lo-mode by computing the WCRT of each task that is released and completed in \lo-mode. 
In \lo-mode, any task may experience \emph{pi-blocking} from accelerator-required tasks in $lpH(\tau_i)$ and $lpL(\tau_i)$. 
We use $PB_i^{\loFN}$ and $B_i^{\loFN}$ to respectively represent the maximum \emph{pi-blocking} time and total blocking time task $\tau_i$ may experience.
\item The schedulability of the \hi-mode by computing the WCRT of each \hi-task that is released and completed in \hi-mode. 
In \hi-mode, a \hi-task may encounter \emph{pi-blocking} due to accelerator-required tasks within both $lpH(\tau_i)$ and $lpL(\tau_i)$, as well as \emph{ci-blocking} resulting from accelerator-required tasks across $lpL(\tau_i)$ and $hpL(\tau_i)$.
We use $PB_i^{\hiFN}$, $CB_i^{\hiFN}$ and $B_i^{\hiFN}$ to respectively represent the maximum \emph{pi-blocking} time, \emph{ci-blocking} time and total blocking time that \hi-task $\tau_i$ may experience.
\item The schedulability of the mode transition by computing the WCRT of each \hi-task that is released in either \lo-mode or mode transition and completed in either \hi-mode or mode transition. 
During mode transition, the blocking scenarios encountered are similar to those in \hi-mode.
We use $PB_i^{*}$, $CB_i^{*}$ and $B_i^{*}$ to respectively represent the maximum \emph{pi-blocking} time, \emph{ci-blocking} time and total blocking time that \hi-task $\tau_i$ may experience.
\end{enumerate}

\noindent \textbf{Response time analysis for \lo-mode.}
We consider a task $\tau_i$ released and finished in \lo-mode. 
The task $\tau_i$ may experience at most one instance of \emph{pi-blocking}. 
In our scheduling model, which involves periodic checks by the scheduler, if a higher-priority task arrives but misses the inspection window, it could face additional blocking equivalent to $T_{sr}$ on top of the longest execution time of instructions from accelerator-required tasks in $lpH(\tau_i)$ and $lpL(\tau_i)$. This duration is represented by Eq.~\ref{1}.
\begin{equation}
\small
PB_i^{\loFN}= I\Bigl(F\bigl(lpH(\tau_i) \cup lpL(\tau_i)\bigl)\Bigl) + T_{sr}
\label{1}
\end{equation}

As there is no criticality inversion in \lo-mode, $B_i^{\loFN}$ is equal to $PB_i^{\loFN}$, as shown in Eq.~\ref{2}:
\begin{equation}
\small
B_i^{\loFN} = PB_i^{\loFN}
\label{2}
\end{equation}

In \lo-mode, the response time $R^{\loFN}_i$ for task $\tau_i$ comprises the sum of its blocking time, execution time, preemption overhead, and the time preempted by other tasks, including the scheduler. The preemption overhead\footnote{In fact, a more rigorous analysis should consider $\max(\Upsilon_{\aFN sr}^\sFN+\Upsilon_{\aFN sr}^\rFN, 2\Upsilon_{\cFN sr}^\cFN)$; however, since context switching involving the accelerator necessarily includes the entire SoC’s context switch. Thus in the worst case, the former strictly exceeds the latter.} is $\Upsilon_{\aFN sr}^\sFN+\Upsilon_{\aFN sr}^\rFN$, with the latter accounting for the potential additional time required if a higher-priority task arrives while a preempted task is restoring its context. The time during which the task is preempted by other tasks includes not only the execution time of those tasks but also the overhead of context save and restore (each preemption incurs one context save and one context restore). The $R^{\loFN}_i$ is presented in Eq.~\ref{4}.
\begin{align}
\small
R^{\loFN}_i &= B_i^{\loFN} + C_i^{\loFN} + \Upsilon_{\aFN sr}^\sFN+\Upsilon_{\aFN sr}^\rFN + \left\lceil \frac{R^{\loFN}_i}{T_{sr}} \right\rceil \Upsilon_{\cFN sr} \notag
\\
&+ \sum_{\tau_j \in \overline{F}\bigl(hpH(\tau_i) \cup hpL(\tau_i)\bigl)} \left\lceil \frac{R^{\loFN}_i}{T_j} \right\rceil (2\Upsilon_{\cFN sr}^\cFN+C_j^{\loFN}) \notag
\\
&+ \sum_{\tau_k \in F\bigl(hpH(\tau_i) \cup hpL(\tau_i)\bigl)} \left\lceil \frac{R^{\loFN}_i}{T_k} \right\rceil (\Upsilon_{\aFN sr}^\sFN+\Upsilon_{\aFN sr}^\rFN+C_k^{\loFN})
\label{4}
\end{align}

\noindent\textbf{Response time analysis for \hi-mode.} We consider a \hi-task released and finished in \hi-mode. 
In \hi-mode, for a \hi-task, all blocking caused by \lo-tasks is treated as \emph{ci-blocking}. Therefore, the \emph{pi-blocking} experienced by a \hi-task should only include blocking caused by lower-priority \hi-tasks (Eq.~\ref{5}).
\begin{equation}
\small
PB_i^{\hiFN}=I\Bigl(F\bigl(lpH(\tau_i)\bigl)\Bigl) + T_{sr}
 \label{5}
\end{equation}

For $CB_i^{\hiFN}$, since in \hi-mode, a \lo-task, regardless of its priority, could only start executing when there are no active \hi-tasks (as detailed in Sec. IV), it is necessary to also consider $hpL(\tau_i)$ in the context of \emph{ci-blocking} for the \hi-task $\tau_i$, as shown in Eq.~\ref{6}.
\begin{equation}
\small
CB_i^{\hiFN}=I\Bigl(F\bigl(lpL(\tau_i) \cup hpL(\tau_i)\bigl)\Bigl) + T_{sr}
 \label{6}
\end{equation}

The $B_i^{\hiFN}$ is the maximum of $CB_i^{\hiFN}$ and $PB_i^{\hiFN}$.
\begin{equation}
\small
B_i^{\hiFN} = I\Bigl(F\bigl(lpL(\tau_i) \cup hpL(\tau_i) \cup lpH(\tau_i)\bigl)\Bigl) + T_{sr}
\label{7}
\end{equation}

\begin{figure*}[]
\centering
\subfigure[Measurement-based method.]{
\label{md1} 
\includegraphics[trim=0.5cm 0.3cm 0.2cm 0cm, clip, scale=0.8]{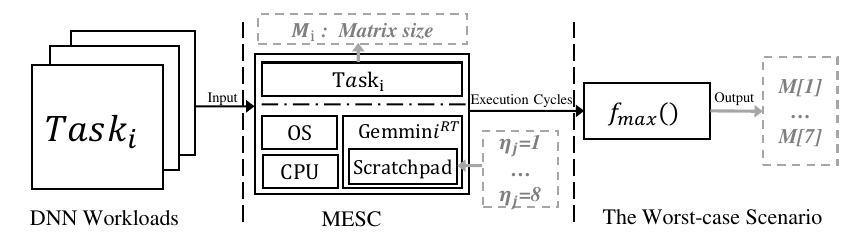} 
}
\subfigure[The worst-case scenario modelling.]{
\label{md2} 
\includegraphics[trim=0.35cm 0.3cm 0.35cm 0.3cm, clip, scale=0.23]{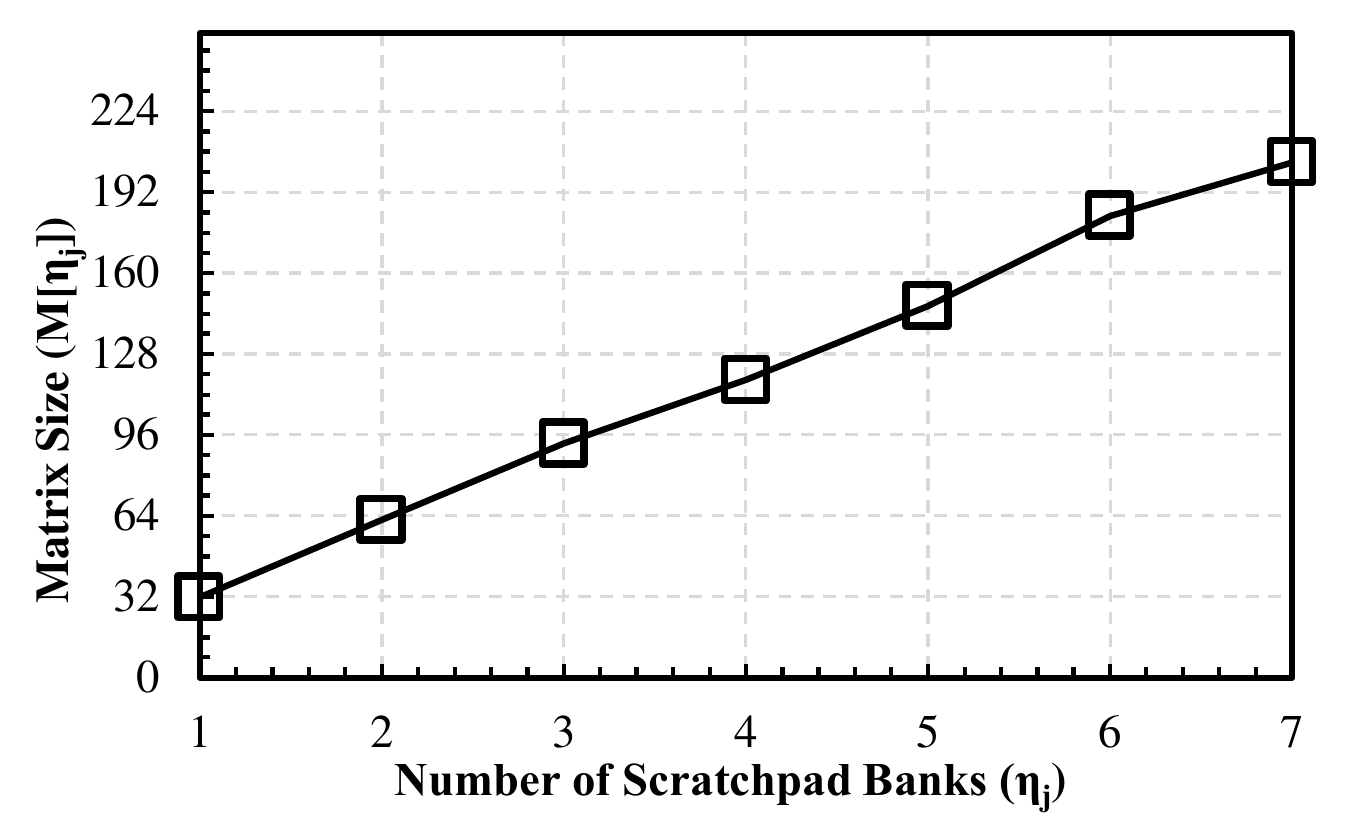} 
}
\caption{Analysis of maximum input matrix sizes accommodated by different scratchpad capacities without impacting task execution speeds, based on varying workloads. The graph plots the sum of the sizes of the original input matrices against scratchpad capacities (the scratchpad consists of 8 banks, each is 32KB).}
\label{md}
\end{figure*}

For \hi-task $\tau_i$, the response time $R^{\hiFN}_i$, as shown in Eq.~\ref{9}, is composed of similar components in \hi-mode as in \lo-mode. Due to the scheduling preference for \hi-tasks in \hi-mode, only tasks from $hpH(\tau_i)$ can preempt $\tau_i$.
\begin{align}
\small
R^{\hiFN}_i &= B_i^{\hiFN} + C_i^{\hiFN} + \Upsilon_{\aFN sr}^\sFN+\Upsilon_{\aFN sr}^\rFN + \left\lceil \frac{R^{\hiFN}_i}{T_{sr}} \right\rceil \Upsilon_{\cFN sr} \notag
\\
&+ \sum_{\tau_j \in \overline{F}\bigl(hpH(\tau_i)\bigl)} \left\lceil \frac{R^{\hiFN}_i}{T_j} \right\rceil (2\Upsilon_{\cFN sr}^\cFN+C_j^{\hiFN}) \notag
\\
&+ \sum_{\tau_k \in F\bigl(hpH(\tau_i)\bigl)} \left\lceil \frac{R^{\hiFN}_i}{T_k} \right\rceil (\Upsilon_{\aFN sr}^\sFN+\Upsilon_{\aFN sr}^\rFN+C_k^{\hiFN})
 \label{9}
\end{align}

\noindent \textbf{Response time analysis for mode transition.}
During the mode transition, for a \hi-task, the experienced \emph{pi-blocking} time, \emph{ci-blocking} time and blocking time are equivalent to those in \hi-mode, as shown in Eq.~\ref{10},~\ref{11} and~\ref{12}.
\begin{equation}
\small
PB_i^{*} = PB_i^{\hiFN}
\label{10}
\end{equation}
\begin{equation}
CB_i^{*} = CB_i^{\hiFN}
\label{11}
\end{equation}
\begin{equation}
B_i^{*} = B_i^{\hiFN}
\label{12}
\end{equation}

Given that any preemption of task $\tau_i$ by \lo-tasks can only occur in \lo-mode, the response time $R^{*}_i$ in mode transition is illustrated in Eq.~\ref{13}.
{\small
\begin{align}
R^{*}_i =& B_i^{*} + C_i^{\hiFN} 
+ \Upsilon_{\aFN sr}^\sFN+\Upsilon_{\aFN sr}^\rFN + \left\lceil \frac{R^{*}_i}{T_{sr}} \right\rceil \Upsilon_{\cFN sr} \notag
\\
+&\sum_{\tau_j \in \overline{F}\bigl(hpL(\tau_i)\bigl)} \left\lceil \frac{R^{\loFN}_i}{T_j} \right\rceil (2\Upsilon_{\cFN sr}^\cFN+C_j^{\loFN}) \notag
\\
+&\sum_{\tau_k \in \overline{F}\bigl(hpH(\tau_i)\bigl)} \left\lceil \frac{R^{*}_i}{T_k} \right\rceil (2\Upsilon_{\cFN sr}^\cFN+C_k^{\hiFN})  \notag
\\
+&\sum_{\tau_m \in F\bigl(hpL(\tau_i)\bigl)} \left\lceil \frac{R^{\loFN}_i}{T_m} \right\rceil (\Upsilon_{\aFN sr}^\sFN+\Upsilon_{\aFN sr}^\rFN+C_m^{\loFN})  \notag
\\
+&\sum_{\tau_n \in F\bigl(hpH(\tau_i)\bigl)} \left\lceil \frac{R^{*}_i}{T_n} \right\rceil (\Upsilon_{\aFN sr}^\sFN+\Upsilon_{\aFN sr}^\rFN+C_n^{\hiFN})
 \label{13}
\end{align}
}

\subsection{Accelerator Local Memory Allocation Method}

To facilitate the work of the address remapper (Sec.~\ref{sec:hardware}), we need an analytical approach to illustrate the relationship between the size of task input matrices and accelerator local memory utilisation.
As pixel density or sequence data increases, the input matrices (or tensors) for DNNs expand, thereby elevating the demand for accelerator local memory and computational requirements. 
This offline analysis helps determine the minimum memory required to maintain task execution speeds without compromising performance.

We illustrate the mathematical relationship between the total size of task original input matrices and the minimal memory required ($\eta_i$ for $\tau_i$) under the constraint of unchanged execution times, as depicted in Fig.~\ref{md1}. Tests were conducted on all tasks mentioned in Sec. III, measuring their original input matrix sizes and execution times across different memory capacities. We then identified the minimal memory size that allows for optimal execution times (i.e., execution times when tasks are not constrained in their use of system resources) without exceeding a set execution threshold, as shown in Fig.~\ref{md2}.
In matrix computations, the generation of a substantial volume of intermediate results typically elevates the demand for accelerator local memory. 
In \gpumodel, however, these intermediate results are stored separately in accumulator, distinct from the primary data storage, which helps mitigate their impact on memory requirements. 
Our modelling is focused on scratchpad. Thus, the correlation between the size of input matrices and the required memory size is remarkably close.

Despite this specific arrangement in \gpumodel, our objective is to establish a universal methodology, not solely applicable to this particular architecture. 
For accelerators where intermediate results are intermixed with the original data, we can apply a similar modelling approach. 
By considering the worst-case scenario, we model both data types together to determine the optimal local memory allocation within the accelerator to ensure efficient processing across different scenarios.

\section{Experimental Evaluation}
\label{sc:EE}

\noindent \textbf{Experimental platform.} We built the \model\ on an AMD Alveo U280 evaluation board, utilising a Rocket core~\cite{asanovic2016rocket} as the RISC-V processor. The processor was instantiated with a 5-stage pipeline and single-width dispatch, featuring a 32 KB L1 I-cache and a 32 KB L1 D-cache, along with a shared 512 KB L2 cache, 4 GB external memory. 
In the \gpumodel\ component, we adhered closely to the default configuration provided by the open-source Gemmini project. This configuration includes a 256 KB scratchpad, a 64 KB accumulator, and a systolic array with a single tile consisting of 256 PEs. 
The system supports a maximum transfer size of 64 bytes with a bus width of 128 for the DMA. The only modification we introduced is an increase in the number of scratchpad banks from 4 to 8.
The software stack, including the OS kernel and workloads, was compiled using the RISC-V GNU toolchain. We adopted FreeRTOS (v.10.5) as the OS, incorporating the modifications detailed in Sec.~\ref{sec:software}.
The system runs at 100 MHz.

\begin{figure*}[t]
\centering
\includegraphics[width=0.92\linewidth]{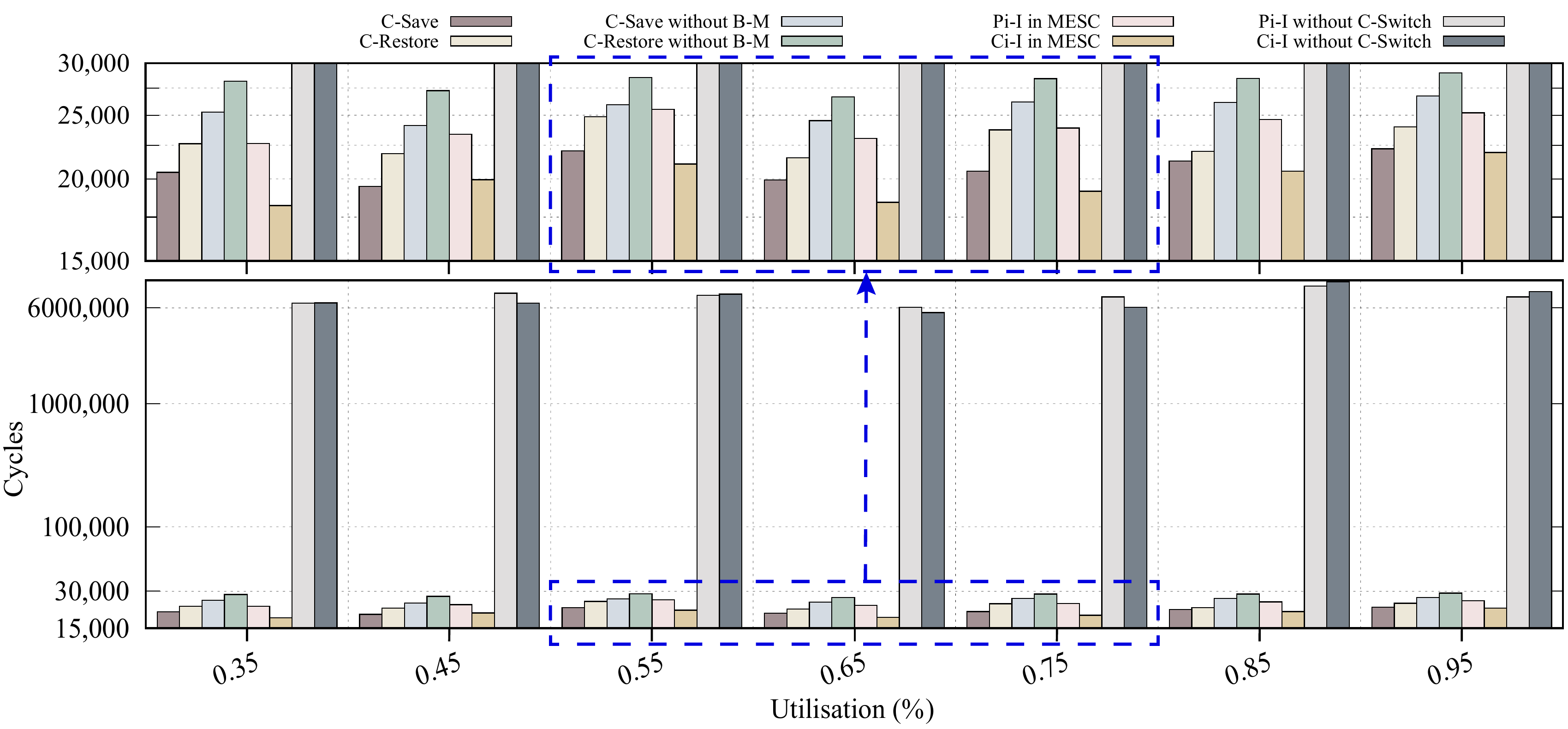}
\caption{Comparison of the impact of different components of \model\ under various utilisations. ``C-Save'' and ``C-Restore'' represent the cycles for context saving and restoring. ``C-Save without B-M'' and ``C-Restore without B-M'' show the cycles without the bank model. ``Pi-I in \modell'' and ``Ci-I in \modell'' denote the cycles for priority inversion and criticality inversion within \model, respectively. ``Pi-I without C-Switch'' and ``Ci-I without C-Switch'' indicate the cycles for priority inversion and criticality inversion without the context-switching mechanism.}
\label{3-2}
\end{figure*}

\noindent \textbf{Task set setup.} To ensure a comprehensive evaluation of \model, we utilise the workloads from the experiments described in Sec.~\ref{sec:motivation} — the DNN workloads and test files developed by UC Berkeley for Gemmini~\cite{genc2019gemmini,genc2021gemmini}. Tasks are randomly selected based on a uniform distribution to form each task set.
The task set parameters used were:
\begin{itemize}
\item Task utilisations $U_i$ were generated using UUnifast~\cite{Bini_2005}, providing an unbiased distribution.
\item The $C_i^{\loFN}$ for each task was directly obtained by retrieving relevant task information from the dataset. The $C_i^{\hiFN}$ was then calculated by $C_i^{\hiFN} = CF \cdot C_i^{\loFN}$, where $CF$ was the criticality factor, default $CF = 2.0$~\cite{burns2020schedulability,bate2022analysis}.
\item The task periods $T_i$ were determined by $T_i = \frac{C_i^{\loFN}}{U_i}$.
\item Tasks were assigned implicit deadlines, implying $D_i = T_i$.
\item Tasks were managed using fixed-priority scheduling, with priorities assigned in descending order of $T_i$.
\item The proportion of \hi-tasks within the task set was determined by the criticality proportion $\gamma$, default $\gamma =0.5$.
\end{itemize}
In our schedulability tests, the total utilisation of the task set was set from $0.5$ to $0.95$, with an incremental step of $0.1$. 
For each utilisation level, $1000$ task sets were generated. 
The number of tasks in each set was determined by $\beta$, default $\beta=10$. 
The operational interval of the scheduler was $T_{sr} = 5000$ cycles.

\subsection{Context-switching Overhead and Blocking Duration}
\noindent \textbf{Experimental setup.} 
We initialised the \model\ in \lo-mode and introduced random disturbances for \hi-tasks by randomly increasing the input of some \hi-tasks before execution, thereby extending their execution times beyond $C_i^{\loFN}$ to prompt mode switch of the system. After running the \model\ for a set duration, we recorded the time taken for context save, context restore, priority inversions, and criticality inversions. Furthermore,  we conducted experiments with the bank allocation strategy and the context-switching mechanism disabled in the \model, aiming to determine the acceleration ratios provided by each component.

\noindent \textbf{Obs. 1.}
By comparing the complete \model\ with the system lacking bank allocations, in terms of context save and restore durations in Fig.~\ref{3-2}, we observe an increase in context-switching times by 4000 to 6000 cycles when the bank allocation is removed. This increased results from the average case, where the address remapper and bank allocation method reduced the frequency and size of data transfers within the accelerator, achieving an acceleration of approximately 20\% to 30\%. 

\noindent \textbf{Obs. 2.}
Comparing the duration of priority inversion and criticality inversion in the complete \model\ in Fig.~\ref{3-2}, we observed that, on average, the duration of criticality inversion was reduced by 3000 to 5000 cycles compared to priority inversion. 
This reduction was due to the system rules in \hi-mode (detailed in Sec.~\ref{sec:approach}). When a \hi-task is preempting a \lo-task, only the data of the currently running \lo-task remains in the scratchpad, significantly increasing the likelihood that the accelerator's local memory is sufficient during the context switch, thus eliminating the need to handle computation data in the scratchpad and then accelerating the context-switching process.
Further comparisons of these durations with a system where context switch functionality was removed revealed significant improvements. For priority inversion durations, we achieved an acceleration exceeding 250 times; for criticality inversion durations, the acceleration surpassed 300 times. This substantial increase in efficiency stems from the fact that without context switch, the accelerator, being a shared resource with long critical-section durations, has to wait until the previous task has completely finished, leading to a substantial waiting period. 
We effectively resolved this issue at the instruction level, allowing timely task transitions and significantly minimising waiting times. 

\begin{figure}[t]
\centering
\includegraphics[width=\linewidth]{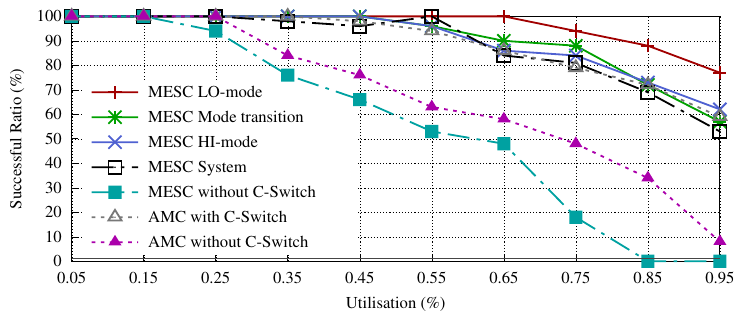}
\caption{Successful ratio of \model\ in different modes, as well as the successful ratio of the entire system with and without context-switching functionality, compared to the successful ratio of the system with and without context switching when applying AMC. ``C-Switch'' denotes context switch.}
\label{1-1}

\end{figure}

\subsection{Successful Ratio}

\noindent \textbf{Experimental setup.} 
We initiated \model\ in \lo-mode and after an initial set running period, recorded the successful ratio (i.e., the proportion of runs with no tasks missing their deadlines during that period) of task sets in \lo-mode, mode transition and \hi-mode. To explore the impact of context switch on successful ratio, we conducted experiments on the \model\ without the context-switching mechanism. We also introduced Adaptive Mixed Criticality (AMC)~\cite{baruah2011response} with a strategy of terminating all \lo-tasks in \hi-mode for comparison. 

\noindent \textbf{Obs. 3.}
Fig.~\ref{1-1} shows the successful ratio of \model\ under different modes in the default conditions. As observed, even at a utilisation rate of 0.95, \model\ maintains considerable schedulability. However, when the context-switching mechanism of \model\ is disabled, its success rate plummets to 0 at a utilisation rate of 0.85. This highlights the crucial impact of the context-switching mechanism on successful ratio.
Additionally, we included comparisons with AMC systems, both with and without context switch. It is evident that without the context-switching mechanism, even terminating all \lo-tasks in \hi-mode does not ensure system success. Furthermore, without the context-switching mechanism, AMC shows significantly higher success rates than \model. However, when both systems incorporate context switch, AMC only marginally outperforms \model\ in scheduling success. This could be attributed to the fact that although \model\ maintains the execution of \lo-tasks in \hi-mode, the context-switching mechanism enables greater flexibility and real-time responsiveness.

\subsection{Successful Ratio in \hi-mode under Various Variables}
\noindent \textbf{Experimental setup.}
Using the same experimental setup as the successful ratio test, we examined the impact of $\gamma$ and $\beta$ on the successful ratio of \hi-tasks in the \hi-mode.

\begin{figure}[t]
\subfigure[Impact of $\gamma$.]{ 
\label{1-2} 
\includegraphics[scale=0.32]{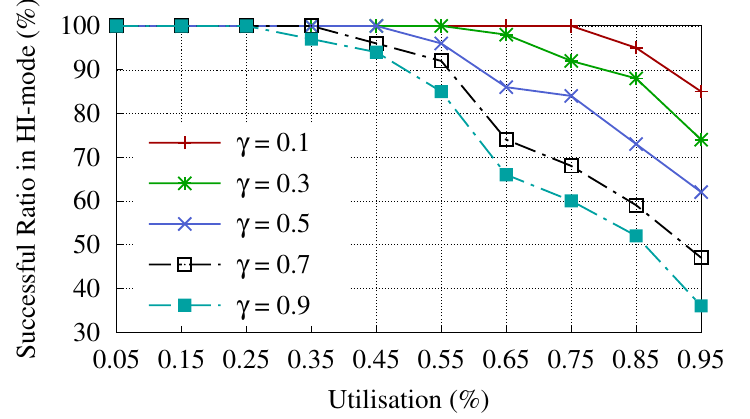} 
}
\subfigure[Impact of $\beta$ .]{ 
\label{1-3} 
\includegraphics[scale=0.32]{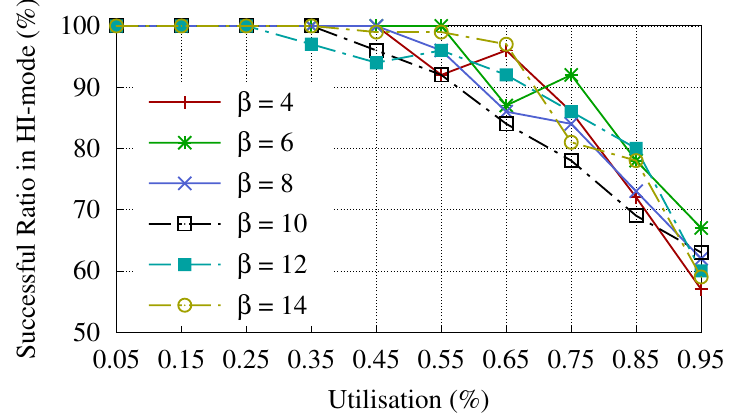} 
}
\caption{Successful ratio of \model\ in \hi-mode under various conditions.}
\label{s1}
\end{figure}

\noindent \textbf{Obs. 4.}
As shown in Fig.~\ref{1-2}, when $\gamma$ increases, the successful ratio of the system significantly decreases in a stable trend. 
This observation is caused by the higher proportion of \hi-tasks, which increases the system load in \hi-mode and intensifies resource contention for the accelerator, thereby reducing the likelihood of all tasks meeting their deadlines.
In contrast, as shown in Fig.~\ref{1-3}, the decrease in the successful ratio is minimal as $\beta$ decreases in \model. 
This is because changes in the number of tasks do not fundamentally increase the system load, and thus do not impact system schedulability as directly as an increase in $\gamma$. Additionally, this stability is partially attributed to the effective context switching, which ensures that the system can promptly handle arriving high-priority tasks regardless of changes in $\beta$, allowing the system to maintain a stable success rate even as the number of tasks decreases.

\subsection{Survivability of \lo-tasks under Various Variables}

We demonstrate the survivability~\cite{jiang2021hiart} of \lo-tasks in \hi-mode by running the workloads mentioned in Sec.~\ref{sec:motivation} using the \model. Survivability is defined as the ratio of the number of \lo-tasks that complete to the number of \lo-tasks released in \hi-mode.

\noindent \textbf{Experimental setup.}
Using the same experimental setup as in the successful ratio test, we examined the impact of $\gamma$ and $\beta$ on the survivability of \lo-tasks in \hi-mode.

\begin{figure}[t]
\subfigure[Impact of $\gamma$.]{ 
\label{2-1} 
\includegraphics[scale=0.32]{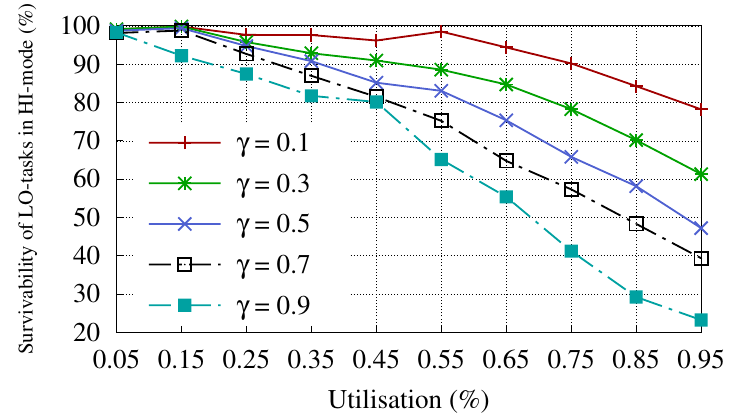} 
}
\subfigure[Impact of $\beta$.]{ 
\label{2-2} 
\includegraphics[scale=0.32]{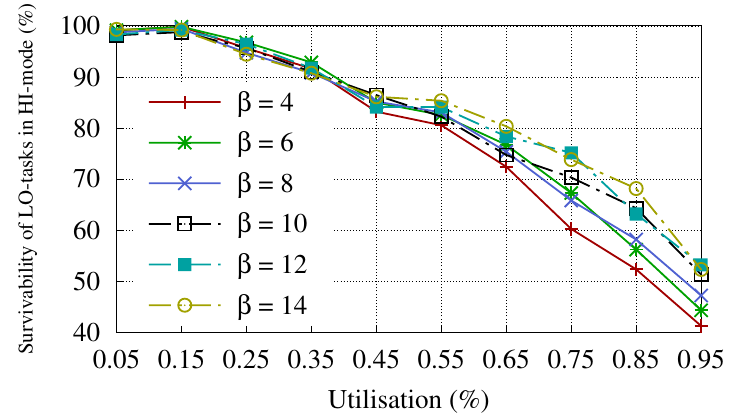} 
}
\caption{Survivability of \lo-tasks in \hi-mode under various conditions.}
\label{s2}
\end{figure}

\noindent \textbf{Obs. 5.}
Fig.~\ref{s2} illustrates the impact of $\gamma$ and $\beta$ on the survivability of \lo-tasks in \hi-mode. As shown in Fig.~\ref{2-1}, the survivability of \lo-tasks significantly decreases with an increase in $\gamma$, as expected. However, even in the most extreme cases, we still maintain over 20\% survivability of \lo-tasks, which is crucial for the stability of the system.
Similar to the successful ratio tests, as shown in Fig.~\ref{2-2}, the impact of $\beta$ on the survivability of \lo-tasks remains very low. This can be attributed to the context-switching mechanism that allows \lo-tasks to run opportunistically, alleviating the pressure on their survivability across different $\beta$, especially at lower utilisation levels.

\subsection{Hardware Overhead}


\noindent \textbf{Experimental setup.}
We compared the hardware overhead and power consumption of \gpu with the default version of Gemmini.
We examined the overhead distribution across its various components, including the controller units, systolic array, reservation station, and scratchpad. The remapping block in the scratchpad is configured as 4KB.
We have synthesised and implemented \gpu on the Alveo U280 FPGA using Vivado (v2021.1) to examine the consumption. The hardware consumption is mainly summarised as the usage of Look-Up-Tables (LUTs) , registers, DSPs, RAMs and power consumption.

\noindent \textbf{Obs. 6.}
The resource efficiency of \gpu is demonstrated in Tbl. II. Note that \gpu only introduced an additional 8815 (4.8\%) LUTs, 2861 (4.3\%) registers, 1 (0.5\%) DSPs, 4KB (1.2\%) RAMs, and 61 (5.2\%) mW. 
Although this comparison is based on Gemmini's open-source default configuration, it is important to note that this configuration includes only a very small systolic array (with a single tile), only 256KB of scratchpad and 64KB of accumulator. In practical scenarios, increasing these configurations would not add to the overhead of the context-switching functionality we introduced, and the relative overhead would likely be even lower.
This is because, even with only a single tile, the overhead and power consumption of the systolic array (64.8\% LUTs, 42.1\% registers, and 45.5\% mW) still constitutes a major portion of the overall hardware costs.

\section{Related Work}

The problems of priority and/or criticality inversions in co-processors have received considerable attention in the research community. Some researchers adopt scheduling protocols to mitigate this problem. For example, GPUsync~\cite{elliott2013gpusync} uses predictive scheduling to allow task status to be transferred among different GPUs.
Similarly, Wu et al.~\cite{wu2021switchflow} optimise resource utilisation by managing the execution of computation graph subgraphs and migrating tasks across different system components. Whilst this scheduling strategy addresses these problems to some extent, it overly relies on hardware resources and struggles to manage scenarios with high task density.
Park et al.~\cite{park2015chimera} propose Chimaera, which abandons the current executing task to free up resources for higher-priority tasks. However, this approach imposes overly strict constraints for real-time application scenarios.
Some researchers have attempted to address these issues at the software level. For instance, Liu et al.~\cite{liu2020removing,liu2023criticality,liu2023criticality1,liu2024taming} partition the software into smaller chunks. Similarly,~\cite{basaran2012supporting,zhou2015gpes,wu2017flep,amert2021timewall} adopt approaches that break tasks into thread-level segments. Although these methods offer improvements, they typically involve limited preemption and necessitate substantial modifications to the software workloads. Hartmann and Margull~\cite{hartmann2019gpuart} introduce a software-based preemption technique that enables controllable preemption at the instruction level, but there is still significant room for improvement in terms of general applicability.
Lee et al.~\cite{lee2020idempotence} implemented fine-grained context switching on GPUs by combining workload partitioning with scheduling, which significantly alleviates the problem. However, this approach incurs a high cost, as it requires rolling back and re-executing previously preempted subtasks.
Some researchers address these issues at the hardware level. Jiang et al.~\cite{jiang2019mcs} propose the MCS-IOV to support context switch for co-processors, but this solution is limited to I/O. Tanasic et al.~\cite{tanasic2014enabling} and Sasongko et al.~\cite{sasongko2021hardware} create hardware enhancements that facilitate preemption by enabling context switch at the thread level. 
Moreover, most industrial embedded GPUs, such as those in the Jetson TX2 (Pascal)~\cite{NVIDIAJetsonTX2,suzen2020benchmark,amert2017gpu}, Xavier (Volta)~\cite{NVIDIAXavier}, and Orin (Ampere)~\cite{NVIDIAOrin} with compute capability below 6.0, also lack the capability for a fine-grained context switch.
Although starting with the Pascal architecture~\cite{NVIDIACUDAGuide} (compute capability 6.0 and higher), these GPUs support instruction-level preemption within compute-type tasks, this granularity still does not fully meet the needs of all tasks requiring a fine-grained context switch.

\begin{table}[]
\centering
\caption{Hardware overhead (implemented on FPGA)}
\includegraphics[width=\linewidth]{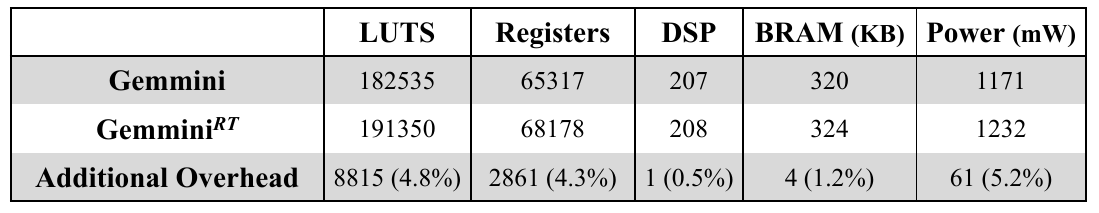}
\label{hwo1}
\end{table}

\begin{table}[]
\centering
\caption{Decomposed overhead of each component in the \gpu}
\includegraphics[width=\linewidth]{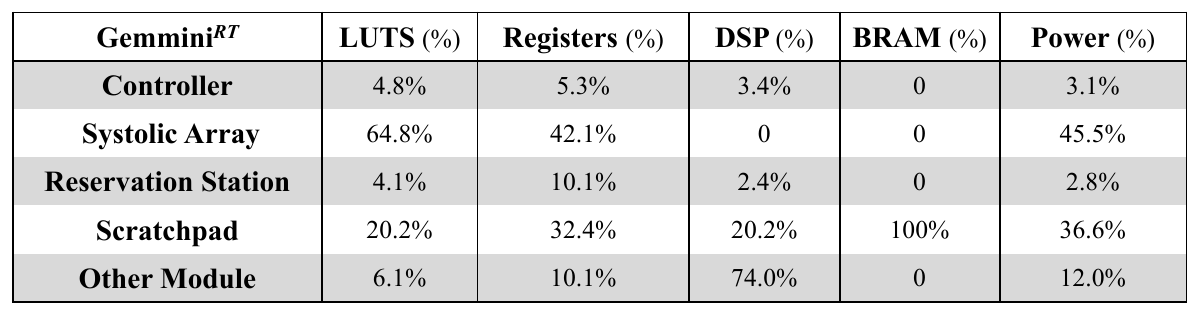}
\label{hwo2}
\end{table}

\section{Conclusion}
In this paper we present MESC, a systematic framework that incorporates hardware and software solutions, forming a comprehensive solution for heterogeneous MCSs. We illustrate how instruction-level context switch for DNN accelerators can be enabled under MESC. 
Through theoretical analysis and experiments, we show that compared to traditional non-preemptive accelerators, the MESC framework achieves an improvement of 2 orders of magnitude in resolving algorithmic priority and criticality inversions. 
Additionally, MESC significantly enhances the system's timing performance and its ability to manage high-priority and \hi-tasks. It also improves the survivability of \lo-tasks in \hi-mode, whilst incurring minimal hardware overhead.

\section{Acknowledgement}
The authors would like to thank the anonymous reviewers and the shepherd for their invaluable feedback on this paper.
This work was supported by the National Natural Science Foundation of China (Grant No. 62472086), the Natural Science Foundation of Jiangsu Province (Grants No. BK20243042), and the Start-up Research Fund of Southeast University (Grant No. RF1028624005). In his first work, Jiapeng Guan would like to express his deepest gratitude to his family, with heartfelt wishes for his father's swift recovery and a healthy life free from pain in the days ahead. Additionally, Jiapeng extends his thanks to Yuanfeng Xu for the supports during system development.


\end{document}